\documentclass[nofootinbib,showkeys,showpacs]{revtex4}
\usepackage[cp1251]{inputenc}
\usepackage[english]{babel}
\usepackage{amsmath}
\usepackage{amssymb}
\usepackage{amsfonts}
\usepackage{latexsym}
\usepackage{dsfont}
\usepackage{bm}
\usepackage{graphics}
\usepackage{empheq}

\newcommand{\nc}{\newcommand}
\newcommand{\rnc}{\renewcommand}

\nc{\be}{\begin{equation}}
\nc{\ee}{\end{equation}}
\nc{\bse}{\begin{equation*}}
\nc{\ese}{\end{equation*}}

\nc{\ba}{\begin{array}}
\nc{\ea}{\end{array}}
\nc{\bea}{\begin{eqnarray}}
\nc{\eea}{\end{eqnarray}}
\nc{\bi}{\begin{itemize}}
\nc{\ei}{\end{itemize}}
\nc{\bn}{\begin{enumerate}}
\nc{\en}{\end{enumerate}}
\nc{\bt}{\begin{tabular}}
\nc{\et}{\end{tabular}}
\nc{\bb}{\begin{equation}\begin{array}{|c|}\hline }
\nc{\eb}{\\ \hline\end{array}\end{equation}}
\nc{\disp}{\displaystyle}

\newcommand{\lb}{\left[}
\newcommand{\rb}{\right]}
\newcommand{\lp}{\left(}
\newcommand{\rp}{\right)}
\newcommand{\lf}{\left\{}
\newcommand{\rf}{\right\}}

\newcommand{\rv}{\right|}
\newcommand{\ld}{\left.}
\newcommand{\rd}{\right.}

\nc{\Def}{\stackrel{def}{=}}
\nc{\sign}{\mathrm{sign}\;}
\nc{\diag}{\mathrm{diag}\;}
\nc{\eq}{\equiv}
\nc{\we}{\wedge}
\nc{\ra}{\rightarrow}
\nc{\bfrac}{\disp\frac}
\nc{\bdel}{\bar{\del}}

\nc{\bfpa}{{\bm{\partial}}}                       
\nc{\pa}[1]{{\partial_{#1}}{}}                    
\nc{\pau}[1]{{\partial^{#1}}{}}                   
\nc{\alp}{\alpha}
\nc{\bet}{\beta}
\nc{\gam}{\gamma}
\nc{\del}{\delta}
\nc{\eps}{\epsilon}
\nc{\veps}{\varepsilon}
\nc{\zet}{\zeta}
\nc{\tet}{\theta}
\nc{\vtet}{\vartheta}
\nc{\iot}{\iota}
\nc{\kap}{\kappa}
\nc{\vkap}{\varkappa}
\nc{\lam}{\lambda}
\nc{\vpi}{\varpi}
\nc{\vrho}{\varrho}
\nc{\sig}{\sigma}
\nc{\vsig}{\varsigma}
\nc{\ups}{\upsilon}
\nc{\vphi}{\varphi}
\nc{\ome}{\omega}

\nc{\Gam}{\Gamma}
\nc{\Del}{\Delta}
\nc{\Tet}{\Theta}
\nc{\Lam}{\Lambda}
\nc{\Sig}{\Sigma}
\nc{\Ups}{\Upsilon}
\nc{\Ome}{\Omega}

\nc{\BL}[1]{\mathbf{#1}}

\nc{\bfa}{\BL{a}}
\nc{\bfb}{\BL{b}}
\nc{\bfc}{\BL{c}}
\nc{\bfd}{\BL{d}}
\nc{\bfe}{\BL{e}}
\nc{\bff}{\BL{f}}
\nc{\bfg}{\BL{g}}
\nc{\bfh}{\BL{h}}
\nc{\bfi}{\BL{i}}
\nc{\bfj}{\BL{j}}
\nc{\bfk}{\BL{k}}
\nc{\bfl}{\BL{l}}
\nc{\bfm}{\BL{m}}
\nc{\bfn}{\BL{n}}
\nc{\bfo}{\BL{o}}
\nc{\bfp}{\BL{p}}
\nc{\bfq}{\BL{q}}
\nc{\bfr}{\BL{r}}
\nc{\bfs}{\BL{s}}
\nc{\bft}{\BL{t}}
\nc{\bfu}{\BL{u}}
\nc{\bfv}{\BL{v}}
\nc{\bfw}{\BL{w}}
\nc{\bfx}{\BL{x}}
\nc{\bfy}{\BL{y}}
\nc{\bfz}{\BL{z}}

\nc{\bfA}{\BL{A}}
\nc{\bfB}{\BL{B}}
\nc{\bfC}{\BL{C}}
\nc{\bfD}{\BL{D}}
\nc{\bfE}{\BL{E}}
\nc{\bfF}{\BL{F}}
\nc{\bfG}{\BL{G}}
\nc{\bfH}{\BL{H}}
\nc{\bfI}{\BL{I}}
\nc{\bfJ}{\BL{J}}
\nc{\bfK}{\BL{K}}
\nc{\bfL}{\BL{L}}
\nc{\bfM}{\BL{M}}
\nc{\bfN}{\BL{N}}
\nc{\bfO}{\BL{O}}
\nc{\bfP}{\BL{P}}
\nc{\bfQ}{\BL{Q}}
\nc{\bfR}{\BL{R}}
\nc{\bfS}{\BL{S}}
\nc{\bfT}{\BL{T}}
\nc{\bfU}{\BL{U}}
\nc{\bfV}{\BL{V}}
\nc{\bfW}{\BL{W}}
\nc{\bfX}{\BL{X}}
\nc{\bfY}{\BL{Y}}
\nc{\bfZ}{\BL{Z}}

\nc{\bfalp}{\bm{\alp}}
\nc{\bfbet}{\bm{\bet}}
\nc{\bfgam}{\bm{\gam}}
\nc{\bfdel}{\bm{\del}}
\nc{\bfeps}{\bm{\eps}}
\nc{\bfveps}{\bm{\veps}}
\nc{\bfzet}{{\bm{\zet}}}
\nc{\bfeta}{\bm{\eta}}
\nc{\bftet}{\bm{\tet}}
\nc{\bfvtet}{\bm{\vtet}}
\nc{\bfiot}{\bm{\iot}}
\nc{\bfkap}{\bm{\kap}}
\nc{\bflam}{\bm{\lam}}
\nc{\bfmu}{\bm{\mu}}
\nc{\bfnu}{\bm{\nu}}
\nc{\bfxi}{\bm{\xi}}
\nc{\bfpi}{\bm{\pi}}
\nc{\bfvpi}{\bm{\vpi}}
\nc{\bfrho}{\bm{\rho}}
\nc{\bfvrho}{\bm{\vrho}}
\nc{\bfsig}{\bm{\sig}}
\nc{\bfvsig}{\bm{\sig}}
\nc{\bftau}{\bm{\tau}}
\nc{\bfups}{\bm{\ups}}
\nc{\bfphi}{\bm{\phi}}
\nc{\bfvphi}{\bm{\vphi}}
\nc{\bfchi}{\bm{\chi}}
\nc{\bfpsi}{\bm{\psi}}
\nc{\bfome}{\bm{\ome}}

\nc{\bfGam}{\bm{\Gam}}
\nc{\bfDel}{\bm{\Del}}
\nc{\bfTet}{\bm{\Tet}}
\nc{\bfLam}{\bm{\Lam}}
\nc{\bfXi}{\bm{\Xi}}
\nc{\bfPi}{\bm{\Pi}}
\nc{\bfSig}{\bm{\Sig}}
\nc{\bfUps}{\bm{\Ups}}
\nc{\bfPhi}{\bm{\Phi}}
\nc{\bfPsi}{\bm{\Psi}}
\nc{\bfOme}{\bm{\Ome}}

\DeclareFontFamily{OT1}{rsfs}{}
\DeclareFontShape{OT1}{rsfs}{m}{n}{<5> rsfs5 <7> rsfs7 <10> rsfs10}{}
\DeclareSymbolFont{mathrsfs}{OT1}{rsfs}{m}{n}
\DeclareSymbolFontAlphabet{\mathrsfs}{mathrsfs}

\nc{\rsA}{\mathrsfs{A}}
\nc{\rsB}{\mathrsfs{B}}
\nc{\rsC}{\mathrsfs{C}}
\nc{\rsD}{\mathrsfs{D}}
\nc{\rsE}{\mathrsfs{E}}
\nc{\rsF}{\mathrsfs{F}}
\nc{\rsG}{\mathrsfs{G}}
\nc{\rsH}{\mathrsfs{H}}
\nc{\rsI}{\mathrsfs{I}}
\nc{\rsJ}{\mathrsfs{J}}
\nc{\rsK}{\mathrsfs{K}}
\nc{\rsL}{\mathrsfs{L}}
\nc{\rsM}{\mathrsfs{M}}
\nc{\rsN}{\mathrsfs{N}}
\nc{\rsO}{\mathrsfs{O}}
\nc{\rsP}{\mathrsfs{P}}
\nc{\rsQ}{\mathrsfs{Q}}
\nc{\rsR}{\mathrsfs{R}}
\nc{\rsS}{\mathrsfs{S}}
\nc{\rsT}{\mathrsfs{T}}
\nc{\rsU}{\mathrsfs{U}}
\nc{\rsV}{\mathrsfs{V}}
\nc{\rsW}{\mathrsfs{W}}
\nc{\rsX}{\mathrsfs{X}}
\nc{\rsY}{\mathrsfs{Y}}
\nc{\rsZ}{\mathrsfs{Z}}

\nc{\CA}[1]{\mathcal{#1}}
\nc{\caA}{\CA{A}}
\nc{\caB}{\CA{B}}
\nc{\caC}{\CA{C}}
\nc{\caD}{\CA{D}}
\nc{\caE}{\CA{E}}
\nc{\caF}{\CA{F}}
\nc{\caG}{\CA{G}}
\nc{\caH}{\CA{H}}
\nc{\caI}{\CA{I}}
\nc{\caJ}{\CA{J}}
\nc{\caK}{\CA{K}}
\nc{\caL}{\CA{L}}
\nc{\caM}{\CA{M}}
\nc{\caN}{\CA{N}}
\nc{\caO}{\CA{O}}
\nc{\caP}{\CA{P}}
\nc{\caQ}{\CA{Q}}
\nc{\caR}{\CA{R}}
\nc{\caS}{\CA{S}}
\nc{\caT}{\CA{T}}
\nc{\caU}{\CA{U}}
\nc{\caV}{\CA{V}}
\nc{\caW}{\CA{W}}
\nc{\caX}{\CA{X}}
\nc{\caY}{\CA{Y}}
\nc{\caZ}{\CA{Z}}

\nc{\lag}{\rsL}      
\nc{\lagG}{\lag^G}   
\nc{\lagM}{\lag^M}   

\nc{\kro}[2]{\del^{#1}_{#2}}

\nc{\bfmet}{\bfg}                     
\nc{\met}[2]{g_{#1 #2}}               
\nc{\metu}[2]{g^{#1 #2}}              
\nc{\rmet}{\sqrt{-g}}                 
\nc{\bfcon}{\bfGam}                   
\nc{\con}[3]{\Gam^{#1}{}_{#2 #3}}     
\nc{\cond}[3]{\Gam_{#1,\, #2 #3}}        
\nc{\bftor}{\bfT}                     
\nc{\tor}[3]{T^{#1}{}_{#2 #3}}        
\nc{\tord}[3]{T_{#1,\, #2 #3}}        
\nc{\toru}[3]{T^{#1,\, #2 #3}}        
\nc{\bfsT}{{\stackrel{*}{\mathbf{T}}}{}}
\nc{\bfstor}{\bfsT}                   
\nc{\sT}{{\stackrel{*}{T}}{}}         
\nc{\stor}[3]{\sT^{#1}{}_{#2 #3}}     
\nc{\stord}[3]{\sT_{#1,\, #2 #3}}     
\nc{\bfctor}{\bfK}                    
\nc{\ctor}[3]{K^{#1}{}_{#2 #3}}       
\nc{\ctord}[3]{K_{#1,\, #2 #3}}       
\nc{\ctoru}[3]{K^{#1,\, #2 #3}}       
\nc{\bfcur}{\bfR}                     
\nc{\cur}[4]{R^{#1}{}_{#2 #3 #4}}     
\nc{\curd}[4]{R_{#1 #2 #3 #4}}        
\nc{\curud}[4]{R^{#1 #2}{}_{#3 #4}}   
\nc{\ric}[2]{R_{#1 #2}}               
\nc{\ricu}[2]{R^{#1 #2}}              
\nc{\bfein}{\bfE}                     
\nc{\ein}[2]{E_{#1 #2}}               
\nc{\einu}[2]{E^{#1 #2}}              
\nc{\bfna}{{\bm{\nabla}}}             
\nc{\na}[1]{{\nabla_{#1}}{}}          
\nc{\nau}[1]{{\nabla^{#1}}{}}         
\nc{\bfsna}{{\stackrel{*}{{\bm{\nabla}}}}{}}         
\nc{\sna}[1]{{\stackrel{*}{\nabla}_{#1}}{}}          
\nc{\snau}[1]{{\stackrel{*}{\nabla}{}^{#1}}{}}       
\nc{\bfdome}{\bfd\bfome\,}                 
\nc{\dome}{dx\,}                           
\nc{\intOme}{\int\limits_{\Ome}}           
\nc{\intb}{\int\limits_{\Sig_1}^{\Sig_2}}  
\nc{\bfdsig}[1]{\bfd\bfsig_{#1}\,}         
\nc{\dsig}[1]{d\sig_{#1}\,}                
\nc{\intdOme}{\oint\limits_{\partial\Ome}} 
\nc{\intSig}{\int\limits_{\Sig}}           
\nc{\intSiga}{\int\limits_{\Sig_1}}        
\nc{\intSigb}{\int\limits_{\Sig_2}}        
\nc{\bfds}[2]{\bfd\bfs_{#1 #2}\,}          
\nc{\ds}[2]{ds_{#1 #2}\,}                  
\nc{\intdSig}{\oint\limits_{\partial\Sig}} 

\nc{\bfdx}[1]{\bfd\bfx^{#1}}          

\nc{\bfgfi}{\bfPhi}                   
\nc{\gfi}[1]{\Phi^{#1}}               
\nc{\gfiA}{\Phi^A}                    
\nc{\gfiB}{\Phi^B}                    
\nc{\bfjfi}{\bfphi}                   
\nc{\jfi}[1]{\phi^{#1}}               
\nc{\jfiA}{\jfi{a}}                    
\nc{\jfiB}{\jfi{b}}                    
\nc{\bfffi}{\bfvphi}                   
\nc{\ffi}[1]{\vphi^{#1}}               
\nc{\ffiA}{\ffi{a}}                    
\nc{\ffiB}{\ffi{b}}                    
\nc{\bfpara}{\bfxi}              
\nc{\para}[1]{\xi^{#1}}          
\nc{\dbfpara}{\del\bfpara}              
\nc{\dpara}[1]{\del\para{#1}}          
\nc{\dbfparK}{\del\bfpara_{\rsK}}              
\nc{\dparK}[1]{\del\para{#1}_{\rsK}}          
\nc{\dx}[1]{\del x^{#1}}
\nc{\dbfLam}{\del\bfLam}
\nc{\dLam}[1]{\del\Lam^{#1}}
\nc{\dSig}{\del\Sig}
\nc{\dIdgfi}[1]{\bfrac{\del I}{\del\gfi{#1}}}
\nc{\dIdgfiA}{\dIdgfi{A}}
\nc{\DIDgfi}[1]{\bfrac{\Del I}{\Del\gfi{#1}}}
\nc{\DIDgfiA}{\DIDgfi{A}}
\nc{\DIDjfi}[1]{\bfrac{\Del I}{\Del\jfi{#1}}}
\nc{\DIDjfiA}{\DIDjfi{a}}
\nc{\DsIDjfi}[1]{\bfrac{\Del^* I}{\Del\jfi{#1}}}
\nc{\DsIDjfiA}{\DsIDjfi{a}}
\nc{\DIDnajfi}[2]{\bfrac{\Del I}{\Del(\na{#1}\jfi{#2})}}
\nc{\DIDnajfiA}[1]{\DIDnajfi{#1}{a}}
\nc{\dIdffi}[1]{\bfrac{\del I}{\del\ffi{#1}}}
\nc{\dIdffiA}{\dIdffi{a}}
\nc{\DIDffi}[1]{\bfrac{\Del I}{\Del\ffi{#1}}}
\nc{\DIDffiA}{\DIDffi{a}}
\nc{\DIDnaffi}[2]{\bfrac{\Del I}{\Del(\na{#1}\ffi{#2})}}
\nc{\DIDnaffiA}[1]{\DIDnaffi{#1}{a}}
\nc{\DIDmet}[2]{\bfrac{\Del I}{\Del \met{#1}{#2}}}
\nc{\DIDtor}[3]{\bfrac{\Del I}{\Del \tor{#1}{#2}{#3}}}
\nc{\DsIDtor}[3]{\bfrac{\Del^* I}{\Del \tor{#1}{#2}{#3}}}
\nc{\DIDnator}[4]{\bfrac{\Del I}{\Del(\na{#1}\tor{#2}{#3}{#4})}}
\nc{\DIDcon}[3]{\bfrac{\Del I}{\Del\con{#1}{#2}{#3}}}
\nc{\DIMDmet}[2]{\bfrac{\Del I^M}{\Del \met{#1}{#2}}}
\nc{\DIMDtor}[3]{\bfrac{\Del I^M}{\Del \tor{#1}{#2}{#3}}}
\nc{\DIMDffi}[1]{\bfrac{\Del I^M}{\Del\ffi{#1}}}
\nc{\DIMDffiA}{\DIMDffi{a}}
\nc{\DIGDmet}[2]{\bfrac{\Del I^G}{\Del \met{#1}{#2}}}
\nc{\DIGDtor}[3]{\bfrac{\Del I^G}{\Del \tor{#1}{#2}{#3}}}

\nc{\K}{K}
\nc{\Kg}[2]{\K^{#1}|_{#2}}
\nc{\KgA}[1]{\Kg{#1}{A}}
\nc{\KgB}[1]{\Kg{#1}{B}}
\nc{\Kj}[2]{\K^{#1}|_{#2}}
\nc{\KjA}[1]{\Kj{#1}{a}}
\nc{\KjB}[1]{\Kj{#1}{b}}
\nc{\sKj}[2]{{}^*\K^{#1}|_{#2}}
\nc{\sKjA}[1]{\sKj{#1}{a}}
\nc{\sKjB}[1]{\sKj{#1}{b}}
\nc{\Kf}[2]{\K^{#1}|_{#2}}
\nc{\KfA}[1]{\Kf{#1}{a}}
\nc{\KfB}[1]{\Kf{#1}{b}}
\nc{\Km}[3]{\K^{#1}|^{#2 #3}}
\nc{\Kt}[4]{\K^{#1}|_{#2}{}^{#3 #4}}
\nc{\sKt}[4]{{}^*\K^{#1}|_{#2}{}^{#3 #4}}
\rnc{\L}{L}
\nc{\Lg}[3]{\L^{#1 #2}|_{#3}}
\nc{\LgA}[2]{\Lg{#1}{#2}{A}}
\nc{\LgB}[2]{\Lg{#1}{#2}{B}}
\nc{\Lj}[3]{\L^{#1 #2}|_{#3}}
\nc{\LjA}[2]{\Lj{#1}{#2}{a}}
\nc{\LjB}[2]{\Lj{#1}{#2}{b}}
\nc{\Lf}[3]{\L^{#1 #2}|_{#3}}
\nc{\LfA}[2]{\Lf{#1}{#2}{a}}
\nc{\LfB}[2]{\Lf{#1}{#2}{b}}
\nc{\Lm}[4]{\L^{#1 #2}|^{#3 #4}}
\nc{\Lt}[5]{\L^{#1 #2}|_{#3}{}^{#4 #5}}
\nc{\gfia}[2]{\gfi{}_{#1}|^{#2}}
\nc{\gfiaA}[1]{\gfia{#1}{A}}
\nc{\gfib}[3]{\gfi{}_{#1}{}^{#2}|^{#3}}
\nc{\gfibA}[2]{\gfib{#1}{#2}{A}}
\nc{\gfic}[4]{\gfi{}_{#1}{}^{#2 #3}|^{#4}}
\nc{\gficA}[3]{\gfic{#1}{#2}{#3}{A}}
\nc{\jfia}[2]{\jfi{}_{#1}|^{#2}}
\nc{\jfiaA}[1]{\jfia{#1}{a}}
\nc{\jfib}[3]{\jfi{}_{#1}{}^{#2}|^{#3}}
\nc{\jfibA}[2]{\jfib{#1}{#2}{a}}
\nc{\jfic}[4]{\jfi{}_{#1}{}^{#2 #3}|^{#4}}
\nc{\jficA}[3]{\jfic{#1}{#2}{#3}{a}}
\nc{\ffia}[2]{\ffi{}_{#1}|^{#2}}
\nc{\ffiaA}[1]{\ffia{#1}{a}}
\nc{\ffib}[3]{\ffi{}_{#1}{}^{#2}|^{#3}}
\nc{\ffibA}[2]{\ffib{#1}{#2}{a}}
\nc{\ffic}[4]{\ffi{}_{#1}{}^{#2 #3}|^{#4}}
\nc{\fficA}[3]{\ffic{#1}{#2}{#3}{a}}

\nc{\bfJpara}{\bfJ[\bfpara]}
\nc{\Jpara}[1]{J^{#1}[\bfpara]}
\nc{\bfJdpara}{\bfJ[\dbfpara]}
\nc{\Jdpara}[1]{J^{#1}[\dbfpara]}
\nc{\bfJdparK}{\bfJ[\dbfparK]}
\nc{\JdparK}[1]{J^{#1}[\dbfparK]}

\nc{\QdxiSig}{Q[\dbfpara;\Sig]}
\nc{\U}[2]{U_{#1}{}^{#2}}
\nc{\Uu}[2]{U^{#1 #2}}
\nc{\Ud}[2]{U_{#1 #2}}
\nc{\M}[3]{M_{#1}{}^{#2 #3}}
\nc{\Mu}[3]{M^{#1 #2 #3}}
\nc{\Md}[3]{M_{#1 #2 #3}}
\nc{\N}[4]{N_{#1}{}^{#2 #3 #4}}
\nc{\Nu}[4]{N^{#1 #2 #3 #4}}
\nc{\Nud}[4]{N^{#1 #2 #3}{}_{#4}}
\nc{\Ia}[1]{I_{#1}}
\nc{\Ib}[2]{I_{#1}{}^{#2}}
\nc{\bfpot}{\bftet}
\nc{\bfpotpara}{\bftet[\bfpara]}
\nc{\potpara}[2]{\tet^{#1 #2}[\bfpara]}
\nc{\bfpotdpara}{\bftet[\dbfpara]}
\nc{\potdpara}[2]{\tet^{#1 #2}[\dbfpara]}
\nc{\bfpotpdpara}{\bftet'[\dbfpara]}
\nc{\potpdpara}[2]{\tet^{'#1 #2}[\dbfpara]}
\nc{\pota}[2]{\tet^{#1 #2}}
\nc{\potb}[3]{\tet_{#1}{}^{#2 #3}}
\nc{\potbu}[3]{\tet^{#1 #2 #3}}
\nc{\potc}[4]{\tet_{#1}{}^{#2 #3 #4}}
\nc{\ppota}[2]{\tet'^{#1 #2}}
\nc{\dpota}[2]{\Del\tet^{#1 #2}}
\nc{\tpotb}[3]{\tilde{\tet}_{#1}{}^{#2 #3}}
\nc{\tpotbu}[3]{\tilde{\tet}^{#1 #2 #3}}
\nc{\tpotc}[4]{\tilde{\tet}_{#1}{}^{#2 #3 #4}}

\nc{\rsJdxi}[1]{\rsJ^{#1}[\dbfpara]}          
\nc{\bfrsJ}{{\bm{\rsJ}}}
\nc{\bfrsJxi}{\bfrsJ[\bfpara]}
\nc{\bfrsJdxi}{\bfrsJ[\dbfpara]}
\nc{\bfJsdpara}{\stackrel{sym}{\bfJ}[\dbfpara]}          
\nc{\Jsdpara}[1]{\stackrel{sym}{J}{}^{#1}[\dbfpara]}          
\nc{\bfrsB}{{\bm{\rsB}}}
\nc{\bfpotBdpara}{\bfrsB[\dbfpara]}
\nc{\potBdpara}[2]{\rsB^{#1 #2}[\dbfpara]}
\nc{\bfpotsdpara}{\stackrel{sym}{\bftet}[\dbfpara]}
\nc{\potsdpara}[2]{\stackrel{sym}{\tet^{#1 #2}}[\dbfpara]}
\nc{\bfUs}{\stackrel{sym}{\bfU}}
\nc{\Us}[2]{\stackrel{sym}{U}{}_{#1}{}^{#2}}
\nc{\bfJsdparK}{\stackrel{sym}{\bfJ}[\dbfparK]}          
\nc{\JsdparK}[1]{\stackrel{sym}{J^{#1}}[\dbfparK]}          

\nc{\A}[3]{A_{#1}{}^{#2 #3}}
\nc{\B}[4]{B_{#1}{}^{#2 #3 #4}}
\nc{\C}[4]{C_{#1}{}^{#2 #3 #4}}
\nc{\youtaa}[2]{\ba{|c|c|}\hline #1 & #2\\ \hline \ea}
\nc{\youtab}[2]{\ba{|c|}\hline 1 \\ \hline 2\\ \hline \ea}
\nc{\youtba}[3]{\ba{|c|c|c|}\hline #1 & #2 & #3\\ \hline \ea}
\nc{\youtbb}[3]{\ba{|c|c|c}\hline #1 & #2\\ \hline #3\\ \cline{1-1} \ea}
\nc{\youtbc}[3]{\ba{|c|}\hline #1 \\ \hline #2\\ \hline #3\\ \hline \ea}
\nc{\yous}[1]{\hat{s}\lp\; #1 \;\rp}
\nc{\youa}[1]{\hat{a}\lp\; #1 \;\rp}
\nc{\Na}[4]{a_{#1}{}^{#2 #3 #4}}
\nc{\bfNb}{\bfb}
\nc{\Nb}[4]{b_{#1}{}^{#2 #3 #4}}
\nc{\bfNc}{\bfc}
\nc{\Nc}[4]{c_{#1}{}^{#2 #3 #4}}
\nc{\Nd}[4]{d_{#1}{}^{#2 #3 #4}}
\nc{\krob}[4]{\del^{{#1} {#2}}_{{#3} {#4}}}
\nc{\Dd}[6]{\Del^{#1 #2 #3}_{\underline{#4 #5 #6}}}
\nc{\Du}[6]{\Del^{\overline{#1 #2 #3}}_{#4 #5 #6}}
\nc{\dbfgfi}{\del\bfgfi}                 
\nc{\dgfi}[1]{\del\gfi{#1}}              
\nc{\dgfiA}{\dgfi{A}}                    
\nc{\dbfjfi}{\del\bfjfi}                 
\nc{\djfi}[1]{\del\jfi{#1}}              
\nc{\djfiA}{\djfi{a}}                    
\nc{\djfiB}{\djfi{b}}                    
\nc{\dbfffi}{\del\bfffi}                 
\nc{\dffi}[1]{\del\ffi{#1}}              
\nc{\dffiA}{\dffi{a}}                    
\nc{\dffib}{\dffi{b}}                    
\nc{\dmet}[2]{\del\met{#1}{#2}}          
\nc{\dmetu}[2]{\del\metu{#1}{#2}}        
\nc{\drmet}{\del\rmet}                   
\nc{\dlag}{\del\lag}                     
\nc{\dcon}[3]{\del\con{#1}{#2}{#3}}      
\nc{\dtor}[3]{\del\tor{#1}{#2}{#3}}      
\nc{\dcur}[4]{\del\cur{#1}{#2}{#3}{#4}}  
\nc{\bdbfgfi}{\bar{\del}\bfPhi}              
\nc{\bdgfiA}{\bar{\del}\Phi^A}               
\nc{\Dbrf}[4]{(\Del^{#1}{}_{#2})\ld^{#3}\rv_{#4}}  
\nc{\DbrfAB}[2]{\Dbrf{#1}{#2}{a}{b}}               
\nc{\Dbrfd}[4]{(\Del_{#1 #2})\ld^{#3}\rv_{#4}}  
\nc{\DbrfdAB}[2]{\Dbrfd{#1}{#2}{a}{b}}               
\nc{\Dbrfu}[4]{(\Del^{#1 #2})\ld^{#3}\rv_{#4}}  
\nc{\DbrfuAB}[2]{\Dbrfu{#1}{#2}{a}{b}}               
\nc{\Dbrj}[4]{(\Del^{#1}{}_{#2})\ld^{#3}\rv_{#4}} 
\nc{\DbrjAB}[2]{\Dbrj{#1}{#2}{a}{b}}              
\nc{\Dbrju}[4]{(\Del^{#1#2})\ld^{#3}\rv_{#4}} 
\nc{\DbrjuAB}[2]{\Dbrju{#1}{#2}{a}{b}}              
\nc{\Dbrjd}[4]{(\Del_{#1#2})\ld^{#3}\rv_{#4}} 
\nc{\DbrjdAB}[2]{\Dbrjd{#1}{#2}{a}{b}}              
\nc{\Dbrg}[4]{(\Del^{#1}{}_{#2})\ld^{#3}\rv_{#4}} 
\nc{\DbrgAB}[2]{\Dbrg{#1}{#2}{A}{B}}              
\nc{\Dbrm}[6]{(\Del^{#1}{}_{#2})\ld_{#3 #4}\rv^{#5 #6}} 
\nc{\Dbrt}[8]{(\Del^{#1}{}_{#2})\ld^{#3}{}_{#4 #5}\rv_{#6}{}^{#7 #8}\,} 
\nc{\Dbrtd}[8]{(\Del_{#1 #2})\ld^{#3}{}_{#4 #5}\rv_{#6}{}^{#7 #8}\,} 
\nc{\Dbrtu}[8]{(\Del^{#1 #2})\ld^{#3}{}_{#4 #5}\rv_{#6}{}^{#7 #8}\,} 
\nc{\Dbrc}[9]{(\Del^{#1}{}_{#2})\ld^{#3}{}_{#4 #5 #6}\rv_{#7}{}^{#8 #9}{}} 

\nc{\dparabfgfi}{\del_{\xi}\bfPhi}    
\nc{\dparagfiA}{\del_{\xi}\Phi^A}     
\nc{\pax}{\partial x}
\nc{\ten}{P}                          
\nc{\meta}[3]{g_{#1}|_{#2 #3}}                
\nc{\metb}[4]{g_{#1}{}^{#2}|_{#3 #4}}         
\nc{\tora}[4]{T_{#1}|^{#2}{}_{#3 #4}}         
\nc{\torb}[5]{T_{#1}{}^{#2}|^{#3}{}_{#4 #5}}  
\nc{\bfnator}{\bfna\bfT}
\nc{\bfnanator}{\bfna\bfna\bfT}
\nc{\bfnaffi}{\bfna\bfvphi}
\nc{\bfnanaffi}{\bfna\bfna\bfvphi}
\nc{\bfnajfi}{\bfna\bfphi}
\nc{\bfnanajfi}{\bfna\bfna\bfphi}
\nc{\dslagdmet}[2]{\bfrac{\partial^*\lag}{\partial\met{#1}{#2}}}
\nc{\dlagdcur}[4]{\bfrac{\partial\lag}{\partial\cur{#1}{#2}{#3}{#4}}}
\nc{\dslagdtor}[3]{\bfrac{\partial^*\lag}{\partial\tor{#1}{#2}{#3}}}
\nc{\dlagdnator}[4]{\bfrac{\partial\lag}{\partial(\na{#1} \tor{#2}{#3}{#4})}}
\nc{\dlagdnanator}[5]{\bfrac{\partial\lag}{\partial(\na{#1}\na{#2} \tor{#3}{#4}{#5})}}
\nc{\dslagdjfi}[1]{\bfrac{\partial^*\lag}{\partial\jfi{#1}}}
\nc{\dslagdjfiA}{\dslagdjfi{a}}
\nc{\dlagdnajfi}[2]{\bfrac{\partial\lag}{\partial(\na{#1}\jfi{#2})}}
\nc{\dlagdnajfiA}[1]{\dlagdnajfi{#1}{a}}
\nc{\dlagdnanajfi}[3]{\bfrac{\partial\lag}{\partial(\na{#1}\na{#2}\jfi{#3})}}
\nc{\dlagdnanajfiA}[2]{\dlagdnanajfi{#1}{#2}{a}}
\nc{\dslagdffi}[1]{\bfrac{\partial^*\lag}{\partial\ffi{#1}}}
\nc{\dslagdffiA}{\dslagdffi{a}}
\nc{\dlagdnaffi}[2]{\bfrac{\partial\lag}{\partial(\na{#1}\ffi{#2})}}
\nc{\dlagdnaffiA}[1]{\dlagdnaffi{#1}{a}}
\nc{\dlagdnanaffi}[3]{\bfrac{\partial\lag}{\partial(\na{#1}\na{#2}\ffi{#3})}}
\nc{\dlagdnanaffiA}[2]{\dlagdnanaffi{#1}{#2}{a}}
\nc{\G}[4]{G_{#1}{}^{#2 #3 #4}}          
\nc{\Gu}[4]{G^{#1 #2 #3 #4}}             
\nc{\Gd}[4]{G_{#1 #2}{}^{#3 #4}}          
\nc{\bfsem}{\bft}
\nc{\sem}[2]{t^{#1}{}_{#2}}
\nc{\semu}[2]{t^{#1 #2}}
\nc{\semd}[2]{t_{#1 #2}}
\nc{\bfsems}{\stackrel{sym}{\bft}}
\nc{\sems}[2]{\stackrel{sym}{t}{}^{#1}{}_{#2}}
\nc{\semsu}[2]{\stackrel{sym}{t}{}^{#1 #2}}
\nc{\semsd}[2]{\stackrel{sym}{t}{}_{#1 #2}}
\nc{\bfsemm}{\stackrel{met}{\bft}}
\nc{\semm}[2]{\stackrel{met}{t}{}^{#1}{}_{#2}}
\nc{\semmu}[2]{\stackrel{met}{t}{}^{#1 #2}}
\nc{\semmd}[2]{\stackrel{met}{t}{}_{#1 #2}}
\nc{\bfsema}{\stackrel{add}{\bft}}
\nc{\sema}[2]{\stackrel{add}{t}{}^{#1}{}_{#2}}
\nc{\semau}[2]{\stackrel{add}{t}{}^{#1 #2}}
\nc{\semad}[2]{\stackrel{add}{t}{}_{#1 #2}}
\nc{\bfsemi}{\stackrel{mod}{\bft}}
\nc{\semi}[2]{\stackrel{mod}{t}{}^{#1}{}_{#2}}
\nc{\semiu}[2]{\stackrel{mod}{t}{}^{#1 #2}}
\nc{\semid}[2]{\stackrel{mod}{t}{}_{#1 #2}}
\nc{\bfsemM}{\bfT}
\nc{\semM}[2]{T^{#1}{}_{#2}}
\nc{\semMu}[2]{T^{#1 #2}}
\nc{\semMd}[2]{T_{#1 #2}}
\nc{\bfsemMs}{\stackrel{sym}{\bfT}}
\nc{\semMs}[2]{\stackrel{sym}{T}{}^{#1}{}_{#2}}
\nc{\semMsu}[2]{\stackrel{sym}{T}{}^{#1 #2}}
\nc{\semMsd}[2]{\stackrel{sym}{T}{}_{#1 #2}}
\nc{\bfsemMm}{\stackrel{met}{\bfT}}
\nc{\semMm}[2]{\stackrel{met}{T}{}^{#1}{}_{#2}}
\nc{\semMmu}[2]{\stackrel{met}{T}{}^{#1 #2}}
\nc{\semMmd}[2]{\stackrel{met}{T}{}_{#1 #2}}
\nc{\bfsemMa}{\stackrel{add}{\bfT}}
\nc{\semMa}[2]{\stackrel{add}{T}{}^{#1}{}_{#2}}
\nc{\semMau}[2]{\stackrel{add}{T}{}^{#1 #2}}
\nc{\semMad}[2]{\stackrel{add}{T}{}_{#1 #2}}
\nc{\bfsemMi}{\stackrel{mod}{\bfT}}
\nc{\semMi}[2]{\stackrel{mod}{T}{}^{#1}{}_{#2}}
\nc{\semMiu}[2]{\stackrel{mod}{T}{}^{#1 #2}}
\nc{\semMid}[2]{\stackrel{mod}{T}{}_{#1 #2}}
\nc{\bfspi}{\bfs}
\nc{\spi}[3]{s^{#1}{}_{#2 #3}}
\nc{\spiu}[3]{s^{#1,\, #2 #3}}
\nc{\spid}[3]{s_{#1,\, #2 #3}}
\nc{\spiud}[3]{s^{#1, #2}{}_{#3}}
\nc{\fj}[3]{{}^{(\jfi{})}f^{#1}{}_{#2 #3}}
\nc{\fju}[3]{{}^{(\jfi{})}f^{#1,\, #2 #3}}
\nc{\fjd}[3]{{}^{(\jfi{})}f_{#1,\, #2 #3}}
\nc{\fjud}[3]{{}^{(\jfi{})}f^{#1,\, #2}{}_{#3}}
\nc{\bfspia}{\stackrel{add}{\bfs}}
\nc{\spia}[3]{\stackrel{add}{s}{}^{#1}{}_{#2 #3}}
\nc{\spiau}[3]{\stackrel{add}{s}{}^{#1,\, #2 #3}}
\nc{\spiad}[3]{\stackrel{add}{s}{}_{#1,\, #2 #3}}
\nc{\spiaud}[3]{\stackrel{add}{s}{}^{#1, #2}{}_{#3}}
\nc{\bfspii}{\stackrel{mod}{\bfs}}
\nc{\spii}[3]{\stackrel{mod}{s}{}^{#1}{}_{#2 #3}}
\nc{\spiiu}[3]{\stackrel{mod}{s}{}^{#1,\, #2 #3}}
\nc{\spiid}[3]{\stackrel{mod}{s}{}_{#1,\, #2 #3}}
\nc{\spiiud}[3]{\stackrel{mod}{s}{}^{#1, #2}{}_{#3}}
\nc{\bfspiM}{\bfS}
\nc{\spiM}[3]{S^{#1}{}_{#2 #3}}
\nc{\spiMu}[3]{S^{#1,\, #2 #3}}
\nc{\spiMd}[3]{S_{#1,\, #2 #3}}
\nc{\spiMud}[3]{S^{#1, #2}{}_{#3}}
\nc{\bfspiMa}{\stackrel{add}{\bfS}}
\nc{\spiMa}[3]{\stackrel{add}{S}{}^{#1}{}_{#2 #3}}
\nc{\spiMau}[3]{\stackrel{add}{S}{}^{#1,\, #2 #3}}
\nc{\spiMad}[3]{\stackrel{add}{S}{}_{#1,\, #2 #3}}
\nc{\spiMaud}[3]{\stackrel{add}{S}{}^{#1, #2}{}_{#3}}
\nc{\bfspiMi}{\stackrel{mod}{\bfS}}
\nc{\spiMi}[3]{\stackrel{mod}{S}{}^{#1}{}_{#2 #3}}
\nc{\spiMiu}[3]{\stackrel{mod}{S}{}^{#1,\, #2 #3}}
\nc{\spiMid}[3]{\stackrel{mod}{S}{}_{#1,\, #2 #3}}
\nc{\spiMiud}[3]{\stackrel{mod}{S}{}^{#1, #2}{}_{#3}}
\nc{\bfbel}{\bfb}
\nc{\bel}[3]{b^{#1}{}_{#2 #3}}
\nc{\belu}[3]{b^{#1 #2 #3}}
\nc{\beld}[3]{b_{#1 #2 #3}}
\nc{\belud}[3]{b^{#1 #2}{}_{#3}}
\nc{\bfbela}{\stackrel{add}{\bfbel}}
\nc{\bela}[3]{\stackrel{add}{b}{}^{#1}{}_{#2 #3}}
\nc{\belau}[3]{\stackrel{add}{b}{}^{#1 #2 #3}}
\nc{\belad}[3]{\stackrel{add}{b}{}_{#1 #2 #3}}
\nc{\belaud}[3]{\stackrel{add}{b}{}^{#1 #2}{}_{#3}}
\nc{\bfbeli}{\stackrel{mod}{\bfbel}}
\nc{\beli}[3]{\stackrel{mod}{b}{}^{#1}{}_{#2 #3}}
\nc{\beliu}[3]{\stackrel{mod}{b}{}^{#1 #2 #3}}
\nc{\belid}[3]{\stackrel{mod}{b}{}_{#1 #2 #3}}
\nc{\beliud}[3]{\stackrel{mod}{b}{}^{#1 #2}{}_{#3}}
\nc{\bfbelM}{\bfB}
\nc{\belM}[3]{B^{#1}{}_{#2 #3}}
\nc{\belMu}[3]{B^{#1 #2 #3}}
\nc{\belMd}[3]{B_{#1 #2 #3}}
\nc{\belMud}[3]{B^{#1 #2}{}_{#3}}
\nc{\bfbelMa}{\stackrel{add}{\bfbelM}}
\nc{\belMa}[3]{\stackrel{add}{B}{}^{#1}{}_{#2 #3}}
\nc{\belMau}[3]{\stackrel{add}{B}{}^{#1 #2 #3}}
\nc{\belMad}[3]{\stackrel{add}{B}{}_{#1 #2 #3}}
\nc{\belMaud}[3]{\stackrel{add}{B}{}^{#1 #2}{}_{#3}}
\nc{\bfbelMi}{\stackrel{mod}{\bfbel}}
\nc{\belMi}[3]{\stackrel{mod}{B}{}^{#1}{}_{#2 #3}}
\nc{\belMiu}[3]{\stackrel{mod}{B}{}^{#1 #2 #3}}
\nc{\belMid}[3]{\stackrel{mod}{B}{}_{#1 #2 #3}}
\nc{\belMiud}[3]{\stackrel{mod}{B}{}^{#1 #2}{}_{#3}}
\nc{\ogG}{\rv_{\lag = \lagG}}
\nc{\ogM}{\rv_{\lag = \lagM}}

\nc{\bfCar}{\rsC}
\nc{\Car}[3]{\rsC^{#1}{}_{#2 #3}}
\nc{\Caru}[3]{\rsC^{#1 #2 #3}}
\nc{\Carud}[3]{\rsC^{#1 #2}{}_{#3}}
\nc{\bfEin}{\rsE}
\nc{\Ein}[2]{\rsE^{#1}{}_{#2}}
\nc{\Einu}[2]{\rsE^{#1 #2}}
\nc{\Eind}[2]{\rsE_{#1 #2}}

\nc{\bfpaffi}{\bfpa\bfffi}
\nc{\bfpapaffi}{\bfpa\bfpa\bfffi}

\nc{\dIdpaffi}[2]{\bfrac{\del I}{\del(\pa{#1}\ffi{#2})}}
\nc{\dIdpaffiA}[1]{\dIdpaffi{#1}{a}}
\nc{\dlagdpaffi}[2]{\bfrac{\partial\lag}{\partial(\pa{#1}\ffi{#2})}}
\nc{\dlagdpaffiA}[1]{\dlagdpaffi{#1}{a}}
\nc{\dlagdpapaffi}[3]{\bfrac{\partial\lag}{\partial(\pa{#1}\pa{#2}\ffi{#3})}}
\nc{\dlagdpapaffiA}[2]{\dlagdpapaffi{#1}{#2}{a}}
\nc{\dIdmet}[2]{\bfrac{\del I}{\del \met{#1}{#2}}}

\nc{\bfMin}{\bfeta}                      
\nc{\Min}[2]{\eta_{#1 #2}}               
\nc{\Minu}[2]{\eta^{#1 #2}}              

\nc{\bftmet}{\tilde{\bfg}}                      
\nc{\tmet}[2]{\tilde{g}{}_{#1 #2}}               
\nc{\tmetu}[2]{\tilde{g}{}^{#1 #2}}              

\nc{\bftna}{\tilde{{\bfna}}}                     
\nc{\tna}[1]{{\tilde{\nabla}{}_{#1}}{}}          
\nc{\tnau}[1]{{\tilde{\nabla}{}^{#1}}{}}         

\nc{\DIDtnaffi}[2]{\bfrac{\Del I}{\Del(\tna{#1}\ffi{#2})}}
\nc{\DIDtnaffiA}[1]{\DIDtnaffi{#1}{a}}
\nc{\dlagdtnaffi}[2]{\bfrac{\partial\lag}{\partial(\tna{#1}\ffi{#2})}}
\nc{\dlagdtnaffiA}[1]{\dlagdtnaffi{#1}{a}}
\nc{\dlagdtnatnaffi}[3]{\bfrac{\partial\lag}{\partial(\tna{#1}\tna{#2}\ffi{#3})}}
\nc{\dlagdtnatnaffiA}[2]{\dlagdtnatnaffi{#1}{#2}{a}}

\nc{\bftcur}{\tilde{\bfR}}                     
\nc{\tcur}[4]{\tilde{R}{}^{#1}{}_{#2 #3 #4}}     
\nc{\tcurd}[4]{\tilde{R}{}_{#1 #2 #3 #4}}        
\nc{\tcurud}[4]{\tilde{R}{}^{#1 #2}{}_{#3 #4}}   
\nc{\bfttor}{\tilde{\bfT}}                     
\nc{\ttor}[3]{\tilde{T}{}^{#1}{}_{#2 #3}}        
\nc{\ttord}[3]{\tilde{T}{}_{#1,\, #2 #3}}        
\nc{\ttoru}[3]{\tilde{T}{}^{#1,\, #2 #3}}        

\nc{\bffsem}{{}^{(\vphi)}\bft}
\nc{\fsem}[2]{{}^{(\vphi)}t^{#1}{}_{#2}}
\nc{\fsemu}[2]{{}^{(\vphi)}t^{#1 #2}}
\nc{\fsemd}[2]{{}^{(\vphi)}t_{#1 #2}}
\nc{\bfmsem}{{}^{(R)}\bft}
\nc{\msem}[2]{{}^{(R)}t^{#1}{}_{#2}}
\nc{\msemu}[2]{{}^{R)}t^{#1 #2}}
\nc{\msemd}[2]{{}^{(R)}t_{#1 #2}}
\nc{\bftsem}{{}^{(T)}\bft}
\nc{\tsem}[2]{{}^{(T)}t^{#1}{}_{#2}}
\nc{\tsemu}[2]{{}^{T)}t^{#1 #2}}
\nc{\tsemd}[2]{{}^{(T)}t_{#1 #2}}
\nc{\bfjsem}{{}^{(\phi)}\bft}
\nc{\jsem}[2]{{}^{(\phi)}t^{#1}{}_{#2}}
\nc{\jsemu}[2]{{}^{(\phi)}t^{#1 #2}}
\nc{\jsemd}[2]{{}^{(\phi)}t_{#1 #2}}
\nc{\bffspi}{{}^{(\vphi)}\bfs}
\nc{\fspi}[3]{{}^{(\vphi)}s^{#1}{}_{#2 #3}}
\nc{\fspiu}[3]{{}^{(\vphi)}s^{#1,\, #2 #3}}
\nc{\fspid}[3]{{}^{(\vphi)}s_{#1,\, #2 #3}}
\nc{\fspiud}[3]{{}^{(\vphi)}s^{#1, #2}{}_{#3}}
\nc{\bfmspi}{{}^{(R)}\bfs}
\nc{\mspi}[3]{{}^{(R)}s^{#1}{}_{#2 #3}}
\nc{\mspiu}[3]{{}^{(R)}s^{#1,\, #2 #3}}
\nc{\mspid}[3]{{}^{(R)}s_{#1,\, #2 #3}}
\nc{\mspiud}[3]{{}^{(R)}s^{#1, #2}{}_{#3}}
\nc{\bftspi}{{}^{(T)}\bfs}
\nc{\tspi}[3]{{}^{(T)}s^{#1}{}_{#2 #3}}
\nc{\tspiu}[3]{{}^{(T)}s^{#1,\, #2 #3}}
\nc{\tspid}[3]{{}^{(T)}s_{#1,\, #2 #3}}
\nc{\tspiud}[3]{{}^{(T)}s^{#1, #2}{}_{#3}}
\nc{\bfjspi}{{}^{(\phi)}\bfs}
\nc{\jspi}[3]{{}^{(\phi)}s^{#1}{}_{#2 #3}}
\nc{\jspiu}[3]{{}^{(\phi)}s^{#1,\, #2 #3}}
\nc{\jspid}[3]{{}^{(\phi)}s_{#1,\, #2 #3}}
\nc{\jspiud}[3]{{}^{(\phi)}s^{#1, #2}{}_{#3}}
\nc{\bffbel}{{}^{(\vphi)}\bfb}
\nc{\fbel}[3]{{}^{(\vphi)}b^{#1}{}_{#2 #3}}
\nc{\fbelu}[3]{{}^{(\vphi)}b^{#1 #2 #3}}
\nc{\fbeld}[3]{{}^{(\vphi)}b_{#1 #2 #3}}
\nc{\fbelud}[3]{{}^{(\vphi)}b^{#1 #2}{}_{#3}}
\nc{\bfmbel}{{}^{(R)}\bfb}
\nc{\mbel}[3]{{}^{(R)}b^{#1}{}_{#2 #3}}
\nc{\mbelu}[3]{{}^{(R)}b^{#1 #2 #3}}
\nc{\mbeld}[3]{{}^{(R)}b_{#1 #2 #3}}
\nc{\mbelud}[3]{{}^{(R)}b^{#1 #2}{}_{#3}}
\nc{\bftbel}{{}^{(T)}\bfb}
\nc{\tbel}[3]{{}^{(T)}b^{#1}{}_{#2 #3}}
\nc{\tbelu}[3]{{}^{(T)}b^{#1 #2 #3}}
\nc{\tbeld}[3]{{}^{(T)}b_{#1 #2 #3}}
\nc{\tbelud}[3]{{}^{(T)}b^{#1 #2}{}_{#3}}
\nc{\bfjbel}{{}^{(\phi)}\bfb}
\nc{\jbel}[3]{{}^{(\phi)}b^{#1}{}_{#2 #3}}
\nc{\jbelu}[3]{{}^{(\phi)}b^{#1 #2 #3}}
\nc{\jbeld}[3]{{}^{(\phi)}b_{#1 #2 #3}}
\nc{\jbelud}[3]{{}^{(\phi)}b^{#1 #2}{}_{#3}}

\begin{document}

%
\title{On the Energy-Momentum and Spin Tensors\\
 in the Riemann-Cartan Space}

\author{Robert R. Lompay}

\affiliation{Department of Physics, Uzhgorod National University\\
54, Voloshyna str., Uzhgorod 88000, Ukraine}

\email{rlompay@gmail.com}

%

\begin{abstract}
General classical theories of material fields in an arbitrary Riemann-Cartan space are considered. For these theories, with the help of equations of balance, new non-trivially generalized, manifestly generally covariant expressions for canonical energy-momentum and spin tensors are constructed in the cases when a Lagrangian contains (a) an arbitrary set of tensorial material fields and their covariant derivatives up to the second order, as well as (b) the curvature tensor and (c) the torsion tensor with its covariant derivatives up to the second order.  A non-trivial manifestly generally covariant generalization of the Belinfante symmetrization procedure, suitable for an arbitrary Riemann-Cartan space, is carried out. A covariant symmetrized energy-momentum tensor is constructed in a general form.


\end{abstract}

\maketitle

%
\section{Introduction}\label{sec_03_00-00}
In the present paper, general classical field theories of an arbitrary set of material fields $\bfffi$ propagating in an arbitrary fixed (external) $(D+1)$-dimensional Riemann-Cartan space $\rsC(1,D)$ are considered. We assume that $\bfffi$ forms a set of tensorial fields with arbitrary but fixed ranks. The aim of the current paper is to construct \emph{general manifestly covariant} expressions for the canonical energy-momentum tensor $\bfsem$ and the spin tensor $\bfspi$ as well as for the symmetrized energy-momentum tensor $\bfsems$ in the cases, when Lagrangian of material fields $\lag$ has the form:
\be\label{sec_03_00-05}
\lag = \lag (\bfmet,\bfcur; \;\bftor,\bfnator,\bfnanator; \;\bfffi,\bfnaffi,\bfnanaffi).
\ee
It depends on an arbitrary set of material fields $\bfffi \Def \{ \ffiA (x) ;\; a=\overline{1,n} \}$ and their first $\bfna\bfffi \Def \{ \na\alp \ffiA (x) \}$ and second $\bfna\bfna\bfffi \Def \{ \na\alp \na\bet \ffiA (x) \}$ covariant derivatives\footnote{We use notations and conventions of the paper~\cite{Lompay_Petrov_2013_a}.}. Additionally, the Lagrangian depends explicitly on the curvature tensor $\bfcur \Def \{ \cur\kap\lam\mu\nu (x) \}$, the torsion tensor $\bftor \Def \{ \tor\lam\mu\nu (x) \}$ and its first  $\bfna\bftor \Def \{ \na\alp \tor\lam\mu\nu (x) \}$  and second $\bfna\bfna\bftor \Def \{ \na\alp \na\bet \tor\lam\mu\nu (x) \}$   covariant derivatives. Thus, the original material Lagrangian contains higher (second) derivatives of the material fields $\bfffi$, and a non-minimal coupling both with the metric $\bfmet \Def \{ \met\mu\nu (x) \}$ (by means of the argument $\bfcur$ in $\lag$) and with the torsion $\bftor$ (by means of the arguments $\bftor$, $\bfnator$ and $\bfnanator$ in $\lag$).

The importance of the stated above task can be seen from the following: The energy momentum tensor (EMT) and the spin tensor (ST) are ones of the most important dynamic characteristics in a field theory both on classical and quantum levels. There are several different types of EMT: canonical $\bfsem \Def \{ \sem\mu\nu (x) \}$, Belinfante symmetrized $\bfsems \Def \{ \sems\mu\nu (x) \}$, metric $\bfsemm \Def \{ \semm\mu\nu (x) \}$ \footnote{Different modifications of these basic EMT both in the Riemann spacetime and in the Riemann-Cartan space were considered in the Refs. \cite{Hehl_1976,Mielke_Hehl_McCrea_1989,Hecht_Hehl_McCrea_Mielke_Neeman_1992,Hannibal_1996}.}. The \emph{metric EMT $\bfsemm$} is the most demanded in the theories defined in a $(D+1)$-dimensional \emph{Riemann spacetime} $\rsR(1,D)$. For a given Lagrangian (and for a well-posed variational problem) the EMT $\bfsemm$ is \emph{uniquely} defined by calculating the variational derivative $\del I/\del\bfmet$ of the action functional $I$ with respect to the metric tensor $\bfmet$ (the Hilbert formula):
\be\label{sec_03_00-06}
\bfrac12 \rmet \semmu\mu\nu  = \dIdmet\mu\nu.
\ee
It is known that whenever the equations of motion $\del I/\del \bfffi = 0$ of material fields $\bfffi$ hold (on the $\bfffi$-equations), the symmetrized EMT $\bfsems$ is equivalent to the metric EMT $\bfsemm$ \cite{Rosenfeld_1940,Rosenfeld_1940_en,Trautman_1966_en,Kopczynski_McCrea_Hehl_1989,Petrov_2008,Lompay_Petrov_2013_b}:
\bse
\sems\mu\nu = \semm\mu\nu\qquad\mbox{(on the $\bfffi$-equations)}.
\ese
Note the canonical spin tensor $\bfspi$ does not appear in the theory explicitly, as it turns out ``hidden'' inside $\bfsems$ (in the form of the Belinfante correction).

Another situation is in metric-torsion theories of gravity. Here, the \emph{canonical EMT $\bfsem$} \cite{Trautman_1972_a,Hehl_1973,Hehl_1974,Hehl_Heyde_Kerlick_Nester_1976,Trautman_2006} turns out to be more useful. However, unlike the metric EMT $\bfsemm$ \eqref{sec_03_00-06}, the canonical EMT $\bfsem$ and ST $\bfspi$, in general, are not so well defined. Indeed, a standard method of construction of these dynamic characteristics in a curved spacetime consists of the following three steps:
\bn
\item In the $(D+1)$-dimensional Minkowski space $\rsM(1,D)$ with a pseudo Cartesian coordinate system (CS) $\{ X^\mu \}$, a Lorentz-invariant field theory defined by an appropriate Lagrangian
\be\label{sec_03_00-15}
\lag = \lag (\bfMin; \; \bfffi, \bfpa\bfffi, \bfpa\bfpa\bfffi, \; \dots),
\ee
is considered, where $\bfMin \Def \lf \Min\mu\nu;\; \mu,\nu = \overline{0,D} \rf = \diag (-1, 1,\dots,1)$ is the metric tensor; $\bfffi \Def \{ \ffiA (X);\; a=\overline{1,n} \}$ is a set of field functions; $\bfpaffi \Def \{\pa\mu \ffiA (X) \}$ and $\bfpapaffi \Def \{\pa\mu \pa\nu \ffiA (X) \}$
are sets of their the first and the second partial derivatives, respectively.

For this Lagrangian by the recipe of the $1$-st Noether theorem the expressions for canonical EMT and ST are constructed:
\be\label{sec_03_00-07}
\sem\mu\nu = \sem\mu\nu (\bfMin; \; \bfffi, \bfpa\bfffi, \bfpa\bfpa\bfffi, \dots);
\ee
\be\label{sec_03_00-08}
\spi\pi\rho\sig = \spi\pi\rho\sig (\bfMin; \; \bfffi, \bfpa\bfffi, \bfpa\bfpa\bfffi, \dots).
\ee
\item By a transition to an arbitrary curved CS $\{ \tilde{x}^\mu \}$ in $\rsM(1,D)$ the expressions \eqref{sec_03_00-07} and \eqref{sec_03_00-08} are transformed into the form
\be\label{sec_03_00-09}
\sem\mu\nu = \sem\mu\nu (\bftmet; \; \bfffi, \bftna\bfffi, \bftna\bftna\bfffi, \dots);
\ee
\be\label{sec_03_00-10}
\spi\pi\rho\sig = \spi\pi\rho\sig (\bftmet; \; \bfffi, \bftna\bfffi, \bftna\bftna\bfffi, \dots),
\ee
where $\bftmet \Def \{ \tmet\mu\nu (\tilde{x}) \}$ and $\bftna \Def \{ \tna\mu \}$ are metric tensor in the space $\rsM(1,D)$ in the CS $\{ \tilde{x}^\mu \}$ and a covariant derivative constructed with its help, respectively.
\item At last, a minimal way of an interaction with gravitational fields is introduced by a formal replacement in the expressions
\be\label{sec_03_00-14}
\tmet\mu\nu\; \ra\; \met\mu\nu, \qquad \tna\mu\; \ra\; \na\mu,
\ee
where  $\bfmet \Def \{ \met\mu\nu (x) \}$ and $\bfna \Def \{ \na\mu \}$ are a metric tensor in a $(D+1)$-dimensional Riemann-Cartan space $\rsC(1,D)$ and a metric-compatible covariant derivative, respectively. Then the expressions \eqref{sec_03_00-09},  \eqref{sec_03_00-10} are transformed into
\be\label{sec_03_00-11}
\sem\mu\nu = \sem\mu\nu (\bfmet; \; \bfffi, \bfna\bfffi, \bfna\bfna\bfffi, \dots);
\ee
\be\label{sec_03_00-12}
\spi\pi\rho\sig = \spi\pi\rho\sig (\bfmet; \; \bfffi, \bfna\bfffi, \bfna\bfna\bfffi, \dots).
\ee
respectively.
\en

The procedure described above gives the unique formula for constructing the canonical EMT and ST in the Riemann-Cartan space $\rsC(1,D)$ (as well as in the Riemann space $\rsR(1,D)$) in the simplest case only, when the original Lagrangian does not contain the derivatives of the fields higher than the first order:
\be
\lag = \lag (\bfMin;\; \bfffi, \bfpa\bfffi).
\ee
 When the scheme \eqref{sec_03_00-15} -- \eqref{sec_03_00-12} is generalized to theories with higher derivatives an ambiguity arises inevitably. It is because a non-commutativity of the covariant derivatives in the space $\rsC(1,D)$ generally takes place
\bse
(\na\mu \na\nu - \na\nu \na\mu )\; \ffiA \Def [\na\mu, \na\nu]\, \ffiA = -\tor\lam\mu\nu \na\lam \ffiA + \cur\kap\lam\mu\nu \DbrfAB\lam\kap \ffiB.
\ese
Here, $\{ \DbrfAB\lam\kap \}$ are the Belinfante-Rosenfeld symbols -- certain combinations of products of the $\del$-Kronecker symbols (for the explicit expressions see, for example, Ref. \cite{Lompay_Petrov_2013_b}).

In the Minkowski space $\rsM(1,D)$, the covariant derivatives commute, therefore the order of the second derivatives in the expressions does not matter, therefore one can write
\be\label{sec_03_00-13}
\tna\mu \tna\nu = \alp \tna\mu \tna\nu + (1-\alp) \tna\nu \tna\mu \qquad (\alp \in \mathds{R}).
\ee
However, an application of the rule \eqref{sec_03_00-14} to the left and the right hand sides of the formula \eqref{sec_03_00-13} gives different results :
\bea
\mbox{L.H.S. of the eq. \eqref{sec_03_00-13}} & \ra & \na\mu \na\nu;\\
\mbox{R.H.S. of the eq. \eqref{sec_03_00-13}} & \ra & \alp \na\mu \na\nu + (1-\alp) \na\nu \na\mu
= \na\nu \na\mu + \alp [ \na\mu, \na\nu ]\nonumber\\
 & = & \na\nu \na\mu + \alp \lp -\tor\lam\mu\nu \na\lam + \cur\kap\lam\mu\nu \Dbrf\lam\kap{}{} \rp.\label{sec_03_02-21}
\eea

On the other hand, it may turn out that due to certain physical requirements, for example, to preserve the conformal invariance \cite{Penrose_1964,Penrose_1964_rep,Chernikov_Tagirov_1968} or the gauge invariance \cite{Itin_Hehl_2003,Bhattacharjee_Chatterjee_2011,Adak_2012}, one should introduce the curvature tensor $\bfcur$ (\emph{non-minimal $\bfmet$-coupling}) and/or the torsion tensor $\bftor$ (\emph{non-minimal $\bftor$-coupling}) into a Lagrangian of a theory in a non-minimal way. Then from the beginning we have the Lagrangian explicitly containing tensors $\bfcur$ and $\bftor$ (and, possibly, their covariant derivatives), that is
\be\label{sec_03_00-16}
\lag = \lag (\bfmet,\bfcur,\; \dots \; ; \;\bftor,\bfnator,\; \dots \; ; \;\bfffi,\bfnaffi,\; \dots).
\ee
One can see that now the procedure \eqref{sec_03_00-15} -- \eqref{sec_03_00-12} cannot be applied because the Lagrangian \eqref{sec_03_00-16} cannot be obtained from the Lagrangian \eqref{sec_03_00-15} by means of the substituion\eqref{sec_03_00-14}. Therefore, the problem of constructing the canonical EMT $\bfsem$ and ST $\bfspi$ in the space $\rsC(1,D)$ when the Lagrangian contains higher derivatives and/or a non-minimal coupling with gravitational fields requires a thorough study.

 In the present paper, to construct the EMT $\bfsem$, $\bfsems$ and ST $\bfspi$, we suggest the following: It is well known that to find the equations of balance, which are satisfied by the tensors $\bfsem$, $\bfsems$ and $\bfspi$ in an arbitrary Riemann-Cartan space, it is enough to know the explicit form of these tensors and the equations of motion of the material fields. However, it is not necessary to know the equations of gravitational fields. Thus, for instance, for the Lagrangian of the form
\be\label{sec_03_00-01}
\lag = \lag (\bfmet; \;\bfffi,\bfnaffi)
\ee
the canonical EMT and ST
\be\label{sec_03_00-03}
\sem\mu\nu = \lag \kro\mu\nu - \dlagdnaffiA\mu \na\nu \ffiA;
\ee
\be\label{sec_03_00-04}
\spi\pi\rho\sig = 2 \dlagdnaffiA\pi \DbrfdAB{[\rho}{\sig]} \ffiB
\ee
satisfy the equation of energy-momentum balance:
\be\label{sec_03_00-02}
\sna\mu \sem\mu\nu = -\sem\mu\lam \tor\lam\mu\nu + \frac12 \spi\pi\rho\sig \curud\rho\sig\pi\nu \qquad \mbox{(on the $\bfffi$-equations)}
\ee
(see, for example, Refs. \cite{Trautman_1972_b,Hehl_1973,Hehl_1974,Hehl_Heyde_Kerlick_Nester_1976,Trautman_2006}).

A converse statement holds as well: If the equations of balance and the equations of motion of material fields are known, this allows to recover expressions for $\bfsem$, $\bfspi$ and $\bfsems$ even in more general cases than \eqref{sec_03_00-01}. To verify this statement we begin from the equation of balance \eqref{sec_03_00-02}, keeping the expressions for $\bfsem$, $\bfspi$ (and for $\bfsems$ as well) as defined and which generalize the expressions \eqref{sec_03_00-03} and \eqref{sec_03_00-04} in case of Lagrangian of the form \eqref{sec_03_00-05}. We will search for these expressions by the method of consistent generalizations, starting from the expressions in the Minkowski space (Sect. \ref{sec_03_01-00}) and, gradually, step by step, passing to more general Lagrangians. Since every such a transition introduce new terms into the expressions for $\bfsem$ and $\bfspi$, we will search for general expressions for the EMT and ST in the form of generalized presentations
\be\label{sec_03_00-17}
\sem\mu\nu = \fsem\mu\nu + \msem\mu\nu + \tsem\mu\nu + \sema\mu\nu;
\ee
\be\label{sec_03_00-18}
\spi\pi\rho\sig = \fspi\pi\rho\sig + \mspi\pi\rho\sig + \tspi\pi\rho\sig + \spia\pi\rho\sig,
\ee
defining consistently items arising in the case of the minimal gravitational coupling in the presence of higher derivatives $\bffsem$, $\bffspi$ (Sect. \ref{sec_03_02-00}); arising in the case of a non-minimal $\bfmet$-coupling $\bfmsem$, $\bfmspi$ (Sect. \ref{sec_03_03-00}), and finally, arising in the case of non-minimal $\bftor$-coupling $\bftsem$, $\bftspi$ and $\bfsema$, $\bfspia$ (Sect. \ref{sec_03_04-00}). In the last section we will find a manifestly generally covariant \emph{non-trivial} generalization of the Belinfante procedure of symmetrization suitable for an arbitrary Riemann-Cartan space. In a general form symmetrized (Belinfante's) EMT $\bfsems$ is constructed, for which a manifestly generally covariant equation of balance is obtained. Some intermediate calculations are carried out in the Appendices \ref{app_03_01-00} -- \ref{app_03_04-00}.

The suggested method allows to define \emph{uniquely} all the terms at the right hand sides of the formulae \eqref{sec_03_00-17} and \eqref{sec_03_00-18}. It is surprisingly that in the term $\bffsem$ (which already appears in the simplest case of minimal gravitational coupling) the second covariant derivatives acquire the \emph{reverse} order in comparison with the original one (see Sect. \ref{sec_03_02-00}, formula \eqref{sec_03_02-19} and the discussion after). Excluding the terms $\bffsem$ and $\bffspi$, all the other terms in the formulae \eqref{sec_03_00-17} and \eqref{sec_03_00-18} as well as the developed generalization of the Belinfante procedure are \emph{new} and did not appear in the literature earlier.

%
\section{The Expressions for the EMT $\bfsem$ and ST $\bfspi$ in the Minkowski Space}\label{sec_03_01-00}

In the Minkowski space with pseudo Cartesian coordinates, consider a classical field-theoretic model described by a Lagrangian of the form
\be\label{sec_03_01-07}
\lag = \lag (\bfffi,\bfpaffi,\bfpapaffi).
\ee
For such a Lagrangian the canonical EMT $\bfsem \Def \lf \sem\mu\nu \rf$ and ST $\bfspi \Def \lf \spi\pi\rho\sig \rf$ constructed by the recipe of the 1-st Noether theorem have the form
\be\label{sec_03_01-05}
\sem\mu\nu = \lag \kro\mu\nu - \dIdpaffiA\mu \pa\nu \ffiA - \dlagdpapaffiA\mu\lam \pa\lam \pa\nu \ffiA;
\ee
\be\label{sec_03_01-06}
\spi\pi\rho\sig = 2 \dIdpaffiA\pi \DbrfdAB{[\rho}{\sig]} \ffiB - 2 \dlagdpapaffiA\pi\lam \lb \Min\lam{[\rho} \pa{\sig]} \ffiA - \DbrfdAB{[\rho}{\sig]} \pa\lam \ffiB \rb.
\ee
Here:
\be
\dIdpaffiA\mu \Def \dlagdpaffiA\mu - \pa\lam \lp \dlagdpapaffiA\lam\mu \rp;
\ee
\be
\DbrfdAB\rho\sig \Def \Min\rho\veps \DbrfAB\veps\sig.
\ee
According to the bracketized indices, antisymmetrization is carried out.

After transition from the pseudo Cartesian coordinates to arbitrary curved ones (in the same Minkowski space $\rsM(1,D)$) the formulae \eqref{sec_03_01-05} and \eqref{sec_03_01-06} acquire the form
\be\label{sec_03_02-06}
\sem\mu\nu = \lag \kro\mu\nu - \DIDtnaffiA\mu \tna\nu \ffiA - \dlagdtnatnaffiA\mu\lam \tna\lam \tna\nu \ffiA;
\ee
\be\label{sec_03_02-20}
\ba{rl}
\spi\pi\rho\sig & = 2 \DIDtnaffiA\pi \DbrfdAB{[\rho}{\sig]} \ffiB\\
& - 2 \dlagdtnatnaffiA\pi\lam \lb \tmet\lam{[\rho} \tna{\sig]} \ffiA - \DbrfdAB{[\rho}{\sig]} \tna\lam \ffiB \rb,
\ea
\ee
where
\be\label{sec_03_02-09}
\DIDtnaffiA\mu \Def \dlagdtnaffiA\mu - \tna\lam \lp \dlagdtnatnaffiA\lam\mu \rp.
\ee

Note that the order of placing the second covariant derivatives in the expressions $\tna\lam \tna\nu \ffiA$ and $\partial\lag/ \partial (\tna\lam \tna\nu \ffiA)$ at this stage is inessential as far as curvature tensor $\bftcur \Def \{ \tcur\kap\lam\mu\nu \}$ and torsion tensor $\bfttor \Def \{ \ttor\lam\mu\nu \}$ of the Minkowski space are equal to zero: $\bftcur = 0$, $\bfttor = 0$. However, in the formulae \eqref{sec_03_02-06} -- \eqref{sec_03_02-09} the order is original, that is in the process of derivation of these relations the order of placing the second derivatives was preserved everywhere. No permutations of derivatives have been done especially.

%
\section{The Generalization to the Case of a Minimal Coupling:\newline the Tensors $\bffsem$ and $\bffspi$}\label{sec_03_02-00}

Let $\rsC(1,D)$ be an arbitrary Riemann-Cartan space and let
\be\label{app_03_01-05}
\lag = \lag (\bfmet; \;\bfffi,\bfnaffi,\bfnanaffi)
\ee
be the Lagrangian, which in the limit of the Minkowski space $\rsM(1,D)$ passes into the Lagrangian \eqref{sec_03_01-07}. Keeping in mind the ambiguity \eqref{sec_03_00-13} -- \eqref{sec_03_02-21}, suppose, that the expressions for $\bffsem \Def \{ \fsem\mu\nu \}$ and $\bffspi \Def \{ \fspi\pi\rho\sig \}$ have the form
\be\label{sec_03_02-05}
\fsem\mu\nu = \lag \kro\mu\nu - \DIDnaffiA\mu \na\nu \ffiA - \dlagdnanaffiA\mu\lam
[ \na\nu \na\lam \ffiA + \alp (\na\lam \na\nu - \na\nu \na\lam) \ffiA];
\ee
\be\label{sec_03_03-06}
\ba{ll}
\fspi\pi\rho\sig & = 2 \DIDnaffiA\pi \DbrjdAB{[\rho}{\sig]} \ffiB\\
& - 2 \dlagdnanaffiA\pi\lam \lb \met\lam{[\rho} \na{\sig]} \ffiA - \DbrjdAB{[\rho}{\sig]} \na\lam \ffiB \rb
\ea
\ee
(compare these formulae with the formulae \eqref{sec_03_02-06} and \eqref{sec_03_02-20}). We notice that in the formula \eqref{sec_03_02-05} the selection of the value $\alp = 0$ leads to the order of placing of the second derivatives which is \emph{reverse} to the original one. The \emph{original} order (as in the formula \eqref{sec_03_02-06}) is reached by the choice $\alp = 1$.

 To find the equation of balance which is satisfied by the EMT $\bffsem$, we calculate its divergence. Using the definition of the modified covariant derivative $\bfsna$
\be
\sna\mu \Def \na\mu + \tor{}{}{}{}_\mu; \qquad \tor{}{}{}{}_\mu \Def \tor\alp\mu\alp,
\ee
the explicit form of EMT $\bffsem$ \eqref{sec_03_02-05} and formula for the calculation of derivative $\na\nu \lag$ \eqref{sec_03_02-07}, after simple calculations we obtain
\be
\ba{l}
\sna\mu \fsem\mu\nu = \tor{}{}{}{}_\nu \lag + \lf \DIDffiA \na\nu \ffiA - \DIDnaffiA\mu [\na\mu, \na\nu] \ffiA \rd\\
\quad \ld - \dlagdnanaffiA\mu\lam [\na\mu, \na\nu] \na\lam \ffiA \rf - \alp \sna\mu \lp \dlagdnanaffiA\mu\lam [\na\lam, \na\nu] \ffiA \rp.
\ea
\ee
In the braces of the last expression, use the formulae for commutator of the covariant derivatives
\bea
[\na\mu, \na\nu] \ffiA & = -\tor\veps\mu\nu \na\veps \ffiA + \cur\sig\rho\mu\nu \DbrjAB\rho\sig \ffiB;
\eea
\bea
[\na\mu, \na\nu] \na\lam \ffiA & = -\tor\veps\mu\nu \na\veps \na\lam \ffiA + \cur\sig\rho\mu\nu \DbrjAB\rho\sig \na\lam \ffiB - \cur\veps\lam\mu\nu \na\veps \ffiA \nonumber\\
 & = -\tor\veps\mu\nu \na\veps \na\lam \ffiA + \cur\sig\rho\mu\nu \lb \DbrjAB\rho\sig \na\lam \ffiB - \kro\rho\lam \na\sig \ffiA \rb.
\eea
Then
\be\label{sec_03_02-10}
\boxed{
\ba{l}
\sna\mu \fsem\mu\nu \eq  - \fsem\mu\lam \tor\lam\mu\nu + \frac12 \fspi\pi\rho\sig \curud\rho\sig\pi\nu + \lf \DIDffiA \na\nu \ffiA \rf\\
\quad - \alp \lf \sna\mu \lp \dlagdnanaffiA\mu\kap [\na\kap, \na\nu] \ffiA \rp + \lp \dlagdnanaffiA\mu\kap [\na\kap, \na\lam] \ffiA \rp \tor\lam\mu\nu \rf.
\ea
}
\ee
At the last stage of deducing the identity \eqref{sec_03_02-10} the formulae \eqref{sec_03_02-05} and \eqref{sec_03_03-06} were also used; the expression $\Del I/ \Del \ffiA$ denotes covariant functional derivative of the action functional $I$ with respect to the material field $\ffiA$:
\bse
\DIDffiA \Def \bfrac{1}{\rmet} \dIdffiA = \dslagdffiA - \sna\mu \lp \dlagdnaffiA\mu \rp + \sna\nu \sna\mu \lp \dlagdnanaffiA\mu\nu \rp.
\ese
Note that the relation \eqref{sec_03_02-10} is just the identity and not the equation as far as in the process of its derivation the equations of motion of the fields were not used.

When the equations of motion $\Del I/\Del \ffiA = 0$ of non-gravitational fields $\bfffi$ hold (\emph{on the $\bfffi$-equations}) the identity \eqref{sec_03_02-10} becomes the \emph{equation of balance}
\be\label{sec_03_02-14}
\boxed{
\ba{l}
\sna\mu \fsem\mu\nu = - \fsem\mu\lam \tor\lam\mu\nu + \bfrac12 \fspi\pi\rho\sig \curud\rho\sig\pi\nu\\
\quad - \alp \lf \sna\mu \lp \dlagdnanaffiA\mu\kap [\na\kap, \na\nu] \ffiA \rp + \lp \dlagdnanaffiA\mu\kap [\na\kap, \na\lam] \ffiA \rp \tor\lam\mu\nu \rf\\
\qquad\qquad\qquad\qquad\qquad \mbox{(on the $\bfffi$-equations).}
\ea
}
\ee
If we require that the equation of balance of the type \eqref{sec_03_00-02} obtained for the theories with Lagrangian of the type \eqref{sec_03_00-01} remains valid in the more general case of the Lagrangian of the type \eqref{app_03_01-05} then  in the formula \eqref{sec_03_02-05} one should choose $\alp = 0$. Thus as a correct generalization of the expressions \eqref{sec_03_00-03} and \eqref{sec_03_00-04} one should derive
\be\label{sec_03_02-19}
\boxed{
\fsem\mu\nu = \lag \kro\mu\nu - \DIDnaffiA\mu \na\nu \ffiA - \dlagdnanaffiA\mu\lam \na\nu \na\lam \ffiA;
}
\ee
\be\label{sec_03_02-22}
\boxed{
\ba{rl}
\fspi\pi\rho\sig & = 2 \DIDnaffiA\pi \DbrfdAB{[\rho}{\sig]} \ffiB\\
 & - 2 \dlagdnanaffiA\pi\lam \lb \met\lam{[\rho} \na{\sig]} \ffiA - \DbrfdAB{[\rho}{\sig]} \na\lam \ffiB \rb.
\ea
}
\ee
We \emph{emphasize} that in the expression $\na\nu \na\lam \ffiA$ in the formula \eqref{sec_03_02-19} the order of placing the second derivatives is \emph{reverse} with respect to the original one (compare with the formula \eqref{sec_03_02-06}). In the case of the \emph{Riemann} spacetime $\rsR(1,D)$, the expressions for the EMT and ST analogous to our expressions \eqref{sec_03_02-19} and \eqref{sec_03_02-22} already have been appeared in the L.~Szabados papers \cite{Szabados_1991,Szabados_1992}. However, no explanations on the reasons of the choice of the order of placing the second derivatives have been given.

In the completion of this section, notice that in the case of a minimal coupling a requirement for the canonical EMT $\bfsem$ to satisfy the equation of balance of the type \eqref{sec_03_00-02} (on the $\bfffi$-equations) fixes form of the tensors $\bfsem$ and $\bfspi$ \emph{quite uniquely}: $\bfsem = \bffsem$,\; $\bfspi = \bffspi$.

%
\section{Generalization to the Case of Non-Minimal $\bfmet$-Coupling:\newline Tensors $\bfmsem$ and $\bfmspi$}\label{sec_03_03-00}

Now, let the Lagrangian $\lag$ explicitly depends on the curvature tensor $\bfcur$, that is
\be\label{sec_03_03-10}
\lag = \lag (\bfmet, \bfcur;\; \bfffi, \bfnaffi, \bfnanaffi).
\ee
In this case choosing EMT $\bffsem$ and ST $\bffspi$ according to the definitions \eqref{sec_03_02-19} and \eqref{sec_03_02-22} and acting as in Sect. \ref{sec_03_02-00} (however, using the identity \eqref{sec_03_02-08} instead of the identity \eqref{sec_03_02-07} herewith), we obtain the identity
\be\label{sec_03_03-05}
\boxed{
\ba{rl}
\sna\mu \fsem\mu\nu & \eq  - \fsem\mu\lam \tor\lam\mu\nu + \frac12 \fspi\pi\rho\sig \curud\rho\sig\pi\nu\\
& + \lf \frac12 \G\alp\bet\gam\del \na\nu \cur\alp\bet\gam\del \rf  + \lf \DIDffiA \na\nu \ffiA \rf.
\ea
}
\ee
Here, $\{ \G\alp\bet\gam\del \} \Def \{ 2 \pa{}\lag / \pa{}\cur\alp\bet\gam\del \}$. It is evidently that the term $\frac12 \Gu\alp\bet\gam\del \na\nu \curd\alp\bet\gam\del$ at the right hand side of the formula \eqref{sec_03_03-05} appears in the case only, when the Lagrangian $\lag$ explicitly depends on the curvature tensor $\bfcur$. One may suppose that this term displays the availability of additional with respect to $\bffsem$ \eqref{sec_03_02-19} and $\bffspi$ \eqref{sec_03_02-22} contributions $\bfmsem$ and $\bfmspi$ to the total EMT $\bfsem$ and ST $\bfspi$ appearing due to the interaction of the fields with the curvature. Then the identity
\be\label{sec_03_03-07}
\ba{rl}
\bfrac12 \Gu\alp\bet\gam\del \na\nu \curd\alp\bet\gam\del & = \lb \sna\mu \lp \Gu\alp\bet\gam\mu \curd\alp\bet\gam\nu \rp + \lp \Gu\alp\bet\gam\mu \curd\alp\bet\gam\lam \rp \tor\lam\mu\nu \rb\\
 & + \bfrac12 \lb (-2)\lp \sna\eta \Gd\rho\sig\pi\eta + \bfrac12 \Gd\rho\sig\veps\eta \tor\pi\veps\eta \rp \rb \curud\rho\sig\pi\nu,
\ea
\ee
proved in the Appendix~\ref{app_03_03-00} allows to define tensors
\be\label{sec_03_03-11}
\boxed{
\msem\mu\nu \Def -\Gu\alp\bet\gam\mu \curd\alp\bet\gam\nu;
}
\ee
\be\label{sec_03_03-08}
\boxed{
\mspi\pi\rho\sig \Def (-2)\lp \sna\eta \Gd\rho\sig\pi\eta + \frac12 \Gd\rho\sig\veps\eta \tor\pi\veps\eta \rp.
}
\ee

Using the formulae \eqref{sec_03_03-07} -- \eqref{sec_03_03-08} we can represent the identity \eqref{sec_03_03-05} in the form
\be\label{sec_03_03-09}
\ba{rl}
\sna\mu \lp \fsem\mu\nu + \msem\mu\nu \rp & \eq - \lp \fsem\mu\lam + \msem\mu\lam \rp \tor\lam\mu\nu\\
 & + \bfrac12 \lp \fspi\pi\rho\sig + \mspi\pi\rho\sig \rp \curud\rho\sig\pi\nu + \DIDffiA \na\nu \ffiA.
\ea
\ee
Hence, we obtain the equation of balance
\be
\ba{rl}
\sna\mu \lp \fsem\mu\nu + \msem\mu\nu \rp & = - \lp \fsem\mu\lam + \msem\mu\lam \rp \tor\lam\mu\nu\\
& + \bfrac12 \lp \fspi\pi\rho\sig + \mspi\pi\rho\sig \rp \curud\rho\sig\pi\nu \quad\mbox{(on the $\bfffi$-equations)}.
\ea
\ee
It is clear that this equation will coincide with the equation \eqref{sec_03_00-02} if in the case of the theories with the non-minimal $\bfmet$-coupling \eqref{sec_03_03-10} we define the canonical EMT $\bfsem$ and ST $\bfspi$ as
\be\label{sec_03_03-12}
\sem\mu\nu = \fsem\mu\nu + \msem\mu\nu;
\ee
\be\label{sec_03_03-13}
\spi\pi\rho\sig = \fspi\pi\rho\sig + \mspi\pi\rho\sig.
\ee
Recall that the quantities presented at the right side of the formulae \eqref{sec_03_03-12} and \eqref{sec_03_03-13} are determined by the definitions \eqref{sec_03_02-19}, \eqref{sec_03_02-22}, \eqref{sec_03_03-11} and \eqref{sec_03_03-08}.

Thus, both in the case of minimal coupling and in the case of non-minimal $\bfmet$-coupling, the requirement, that the canonical EMT $\bfsem$ must satisfy the equation of balance of the type \eqref{sec_03_00-02} (on the $\bfffi$-equations) fixes form of the tensors $\bfsem$ and $\bfspi$ \emph{uniquely}.

%
\section{Generalization to the Case of a Non-Minimal $\bftor$-Coupling:\newline Tensors $\bftsem$, $\bftspi$ and $\bfsema$, $\bfspia$}\label{sec_03_04-00}
At last, consider the case when in addition the Lagrangian $\lag$ explicitly depends on the torsion tensor $\bftor$, its first $\bfnator$ and second $\bfnanator$ covariant derivatives, that is
\be\label{sec_03_04-25}
\lag = \lag (\bfmet, \bfcur;\; \bftor, \bfnator,\bfnanator;\; \bfffi, \bfnaffi, \bfnanaffi).
\ee

\subsection{The Torsion Field $\bftor$ as a Physical Field. Tensors $\bftsem$ and $\bftspi$}

The explicit dependence of the Lagrangian \eqref{sec_03_04-25} on the torsion field $\bftor$ makes this field similar to the usual mater fields $\bfffi$ propagating in the space-time $\rsC(1,D)$. Therefore, we can expect that total canonical EMT $\bfsem$ and ST $\bfspi$ will contain the contributions $\bftsem$ and $\bftspi$ induced by the field $\bftor$, whose structure is analogous to the structure of the contributions $\bffsem$ and $\bffspi$ induced by the fields $\bfffi$. Thus, by analogy with the formulae \eqref{sec_03_02-19} and \eqref{sec_03_02-22} we choose tensors $\bftsem$ and $\bftspi$ in the form
\be
\boxed{
\tsem\mu\nu =  - \DIDnator\mu\alp\bet\gam \na\nu \tor\alp\bet\gam - \dlagdnanator\mu\lam\alp\bet\gam \na\nu \na\lam \tor\alp\bet\gam;
}
\ee
\be
\boxed{
\ba{rl}
\tspi\pi\rho\sig & = 2 \DIDnator\pi\alp\bet\gam \Dbrtd{[\rho}{\sig]}\alp\bet\gam\eta\zet\xi \tor\eta\zet\xi\\
 & - 2 \dlagdnanator\pi\lam\alp\bet\gam \lb \met\lam{[\rho} \na{\sig]} \tor\alp\bet\gam - \Dbrtd{[\rho}{\sig]}\alp\bet\gam\eta\zet\xi \na\lam \tor\eta\zet\xi \rb.
\ea
}
\ee
Then, acting as in Sects. \ref{sec_03_02-00} and \ref{sec_03_03-00} (however, using now the identity \eqref{sec_03_04-20}), we obtain the identity
\be\label{sec_03_04-21}
\boxed{
\ba{l}
\sna\mu \lp \fsem\mu\nu + \msem\mu\nu + \tsem\mu\nu \rp \eq - \lp \fsem\mu\lam + \msem\mu\lam + \tsem\mu\lam \rp \tor\lam\mu\nu\\
\quad + \bfrac12 \lp \fspi\pi\rho\sig + \mspi\pi\rho\sig + \tspi\pi\rho\sig \rp \curud\rho\sig\pi\nu
+ \DsIDtor\alp\bet\gam \na\nu \tor\alp\bet\gam + \DIDffiA \na\nu \ffiA.
\ea
}
\ee
In the last formula
\be\label{sec_03_04-24}
\DsIDtor\alp\bet\gam \Def \dslagdtor\alp\bet\gam - \sna\mu \lp \dlagdnator\mu\alp\bet\gam \rp + \sna\nu \sna\mu \lp \dlagdnanator\mu\nu\alp\bet\gam \rp.
\ee
For the sake of simplicity we can unite the fields $\bftor$ and $\bfffi$ into the unique set $\bfjfi \Def \{ \jfiA \} \Def \{ \bftor, \bfffi \}$. Further, thus we define the total canonical EMT $\bfsem$ and ST $\bfspi$ as
\be\label{sec_03_04-06}
\boxed{
\sem\mu\nu \Def \fsem\mu\nu + \msem\mu\nu + \tsem\mu\nu = \jsem\mu\nu + \msem\mu\nu;
}
\ee
\be\label{sec_03_04-12}
\boxed{
\spi\pi\rho\sig \Def \fspi\pi\rho\sig + \mspi\pi\rho\sig + \tspi\pi\rho\sig = \jspi\pi\rho\sig + \mspi\pi\rho\sig,
}
\ee
where
\be
\jsem\mu\nu \Def \fsem\mu\nu + \tsem\mu\nu \Def \lag \kro\mu\nu - \DIDnajfiA\mu \na\nu \jfiA - \dlagdnanajfiA\mu\lam \na\nu \na\lam \jfiA;
\ee
\be
\ba{rl}
\jspi\pi\rho\sig \Def \fspi\pi\rho\sig + \tspi\pi\rho\sig
\Def & 2 \DIDnajfiA\pi \DbrjdAB{[\rho}{\sig]} \jfiB - 2 \dlagdnanajfiA\pi\lam\\
& \times\lb \met\lam{[\rho} \na{\sig]} \jfiA - \DbrjdAB{[\rho}{\sig]} \na\lam \jfiB \rb.
\ea
\ee
The expressions for EMT $\bfsem$ \eqref{sec_03_04-06} and ST $\bfspi$ \eqref{sec_03_04-12} include the torsion field $\bftor$ and material fields $\bfffi$ in a \emph{maximum equal way}.

The variational derivative $\Del I/ \Del \tor\alp\bet\gam$ of the action functional $I$ with respect to the torsion field $\tor\alp\bet\gam$ has the following structure \cite{Lompay_Petrov_2013_b}:
\be\label{sec_03_04-22}
\DIDtor\alp\bet\gam = \DsIDtor\alp\bet\gam + \frac12 \belud\gam\bet\alp,
\ee
where
\be\label{sec_03_04-23}
\belu\gam\bet\alp \Def \Du\gam\bet\alp\pi\rho\sig \lp \jspiu\pi\rho\sig + \mspiu\pi\rho\sig \rp = \Du\gam\bet\alp\pi\rho\sig \spiu\pi\rho\sig
\ee
is the Belinfante tensor, induced by the ST $\bfspi$ and
\bse
\Du\alp\bet\gam\pi\rho\sig \Def \frac{1}{2} \lp \kro\bet\pi \kro\alp\rho \kro\gam\sig + \kro\gam\pi \kro\alp\rho \kro\bet\sig - \kro\alp\pi \kro\bet\rho \kro\gam\sig \rp.
\ese
Therefore, the last two items in the identity \eqref{sec_03_04-21} can be represented equivalently as
\be\label{sec_03_04-07}
\ba{l}
\DsIDtor\alp\bet\gam \na\nu \tor\alp\bet\gam + \DIDffiA \na\nu \ffiA
=  -\bfrac12 \belud\gam\bet\alp \na\nu \tor\alp\bet\gam\\
\quad + \DIDtor\alp\bet\gam \na\nu \tor\alp\bet\gam + \DIDffiA \na\nu \ffiA \Def -\bfrac12 \belud\gam\bet\alp \na\nu \tor\alp\bet\gam + \DIDjfiA \na\nu \jfiA.
\ea
\ee
Taking into account the formulae \eqref{sec_03_04-06}, \eqref{sec_03_04-12} and \eqref{sec_03_04-07}, the identity \eqref{sec_03_04-21} acquires the form
\be\label{sec_03_04-19}
\boxed{
\sna\mu \sem\mu\nu \eq - \sem\mu\lam \tor\lam\mu\nu + \bfrac12 \spi\pi\rho\sig \curud\rho\sig\pi\nu - \bfrac12 \belud\gam\bet\alp \na\nu \tor\alp\bet\gam + \DIDjfiA \na\nu \jfiA.
}
\ee
Hence, on the $\bfjfi$-equations (that is on the $\bftor$- \emph{and} $\bfffi$-equations) we get the equation of balance
\be\label{sec_03_04-28}
\boxed{
\sna\mu \sem\mu\nu = - \sem\mu\lam \tor\lam\mu\nu + \bfrac12 \spi\pi\rho\sig \curud\rho\sig\pi\nu - \bfrac12 \belud\gam\bet\alp \na\nu \tor\alp\bet\gam \quad\mbox{(on the $\bfjfi$-equations)}.
}
\ee
In comparison with the equation \eqref{sec_03_00-02}, the last equation contains at the right hand side an \emph{additional} item $\{ - \bfrac12 \belud\gam\bet\alp \na\nu \tor\alp\bet\gam \}$ and is valid on the $\bfjfi$-equations (not on the $\bfffi$--equations). This is the price, which must be paid for a desire to consider the torsion field $\bftor$ as a physical field, like the material fields $\bfffi$. Although the presence of the additional terms in the constructed EMT $\bfsem$ \eqref{sec_03_04-06} and ST $\bfspi$ \eqref{sec_03_04-12} looks at as contradicting to the standard definitions, it is necessary.  Indeed, in the general metric-torsion theories of gravitation conserved gravitational current cannot be  constructed without these terms \cite{Lompay_Petrov_2013_b}.

\subsection{Generalization of the Belinfante Symmetrization Procedure}
Let us consider a problem of constructing the symmetrized EMT $\bfsems$ basing on the canonical EMT $\bfsem$ \eqref{sec_03_04-06} and ST $\bfspi$ \eqref{sec_03_04-12}. Using in the formula \eqref{sec_03_04-19} the identity \eqref{sec_03_04-08} we can represent it in the form
\be\label{sec_03_04-09}
\ba{l}
\sna\mu \lb \sem\mu\nu + \lp \sna\eta \belud\mu\eta\nu + \bfrac12 \belud\veps\eta\nu \tor\mu\veps\eta + \belud\mu\bet\alp \tor\alp\bet\nu \rp \rb\\
\eq -\lb \sem\mu\lam + \lp \sna\eta \belud\mu\eta\lam + \bfrac12 \belud\veps\eta\lam \tor\mu\veps\eta + \belud\mu\bet\alp \tor\alp\bet\lam \rp \rb \tor\lam\mu\nu + \DIDjfiA\na\nu \jfiA.
\ea
\ee
In the last formula a combination of the Belinfante tensor $\bfbel \Def \{ \belu\gam\bet\alp \}$ in the parentheses represents generalized (for the case of presence of a torsion) the Belinfante correction \cite{Belinfante_1939,Belinfante_1940}. Therefore, if we define the \emph{symmetrized EMT} $\bfsems$  as%
\be\label{sec_03_04-16}
\boxed{
\sems\mu\nu \Def \sem\mu\nu + \lp \sna\eta \belud\mu\eta\nu + \bfrac12 \belud\veps\eta\nu \tor\mu\veps\eta + \belud\mu\bet\alp \tor\alp\bet\nu \rp,
}
\ee
the identity \eqref{sec_03_04-09} turns into:
\be
\boxed{
\sna\mu \sems\mu\nu \eq -\sems\mu\lam \tor\lam\mu\nu + \DIDjfiA \na\nu \jfiA.
}
\ee
On the $\bfjfi$-equations we have a corespondent equation of balance:
\be\label{sec_03_04-10}
\boxed{
\sna\mu \sems\mu\nu = -\sems\mu\lam \tor\lam\mu\nu \qquad \mbox{(on the $\bfjfi$-equations)}.
}
\ee
As it should be for the symmetrized EMT $\bfsems$, spin tensor $\bfspi$ does not enter the formula \eqref{sec_03_04-10}.

\subsection{The Torsion Field $\bftor$ as a Geometrical Field. Tensors $\bfsema$ and $\bfspia$}

Now, recall that in the field theories defined in the Riemann-Cartan space $\rsC(1,D)$, the torsion field $\bftor$ enters the Lagrangian not only explicitly (through the arguments $\bftor$, $\bfnator$ and $\bfnanator$), but also \emph{implicitly}. In fact, the torsion $\bftor$ is included in the \emph{geometrical structure} of space-time $\rsC(1,D)$ itself. Indeed, the torsion $\bftor$ enters a connection $\bfcon \Def \{ \con\lam\mu\nu(x) \}$, with the use of which both the covariant derivatives $\bfna$ and the curvature tensor $\bfcur$ are constructed. Thus, there are no reasons to expect that search for expressions for the total EMT $\bfsem$ and ST $\bfspi$ contain contributions from the torsion field $\bftor$ and material fields $\bfffi$ in completely equal way. From such a point of view the presence of the additional term $(-\frac12 \belud\gam\bet\alp \na\nu \tor\alp\bet\gam)$ in the equation \eqref{sec_03_04-28} can be treated as an indication to the existence in the total EMT and ST \emph{additional} with respect to $\bftsem$ and $\bftspi$ terms, which we denote as $\bfsema$ and $\bfspia$. Evidently, such terms destroy a formal similarity between $\bftor$ and $\bfffi$ as \emph{physical} fields. To find these terms note the following. The structure of the variational derivative $\{ \Del I/ \Del \tor\alp\bet\gam \}$ \eqref{sec_03_04-22}, \eqref{sec_03_04-23} shows that it is necessary to introduce an additional ST $\bfspia = \{ \spia\pi{[\rho}{\sig]} = \spia\pi\rho\sig \}$ and the Belinafante tensor $\bfbela = \{ \belau\gam\bet\alp \}$, induced by it in the way that the relationships
\be\label{sec_03_04-26}
\frac{1}{2} \belaud\gam\bet\veps = \DsIDtor\veps\bet\gam.
\ee
and
\be
\belau\gam\bet\alp \Def \Du\gam\bet\alp\pi\rho\sig \spiau\pi\rho\sig
\ee
take a place. By the last two formulae, it is easy to determine an explicit form of the tensor $\bfspia$:
\be
\boxed{
\spiau\pi\rho\sig = -4 \metu{[\sig|}\veps \DsIDtor\veps{|\rho]}\pi.
}
\ee
Then, it is evidently,
\be\label{sec_03_04-27}
\DIDtor\alp\bet\gam = \frac12 \lp \fbelud\gam\bet\alp + \mbelud\gam\bet\alp + \tbelud\gam\bet\alp + \belaud\gam\bet\alp \rp,
\ee
where
\be
\ba{c}
\fbelu\gam\bet\alp \Def \Du\gam\bet\alp\pi\rho\sig \fspiu\pi\rho\sig, \qquad
\mbelu\gam\bet\alp \Def \Du\gam\bet\alp\pi\rho\sig \mspiu\pi\rho\sig,\\
\tbelu\gam\bet\alp \Def \Du\gam\bet\alp\pi\rho\sig \tspiu\pi\rho\sig.
\ea
\ee

Let us return to the identity \eqref{sec_03_04-21}. Using the formula \eqref{sec_03_04-08} with the exchange $\belud\gam\bet\alp = \belaud\gam\bet\alp$, transform the term
\bse
\DsIDtor\alp\bet\gam \na\nu \tor\alp\bet\gam = \frac12 \belaud\gam\bet\alp \na\nu \tor\alp\bet\gam.
\ese
Then the identity takes the form
\be\label{sec_03_04-11}
\ba{l}
\sna\mu \lb \fsem\mu\nu + \msem\mu\nu + \tsem\mu\nu - \lp \sna\eta \belaud\mu\eta\nu + \bfrac12 \belaud\veps\eta\nu \tor\mu\veps\eta + \belaud\mu\bet\alp \tor\alp\bet\nu \rp \rb \\
\eq \lb \fsem\mu\lam + \msem\mu\lam + \tsem\mu\lam - \lp \sna\eta \belaud\mu\eta\lam + \bfrac12 \belaud\veps\eta\lam \tor\mu\veps\eta + \belaud\mu\bet\alp \tor\alp\bet\lam \rp \rb\\
\times(-\tor\lam\mu\nu) + \bfrac12 \lp \fspi\pi\rho\sig + \mspi\pi\rho\sig + \tspi\pi\rho\sig + \spia\pi\rho\sig \rp \curud\rho\sig\pi\nu + \DIDffiA \na\nu \ffiA.
\ea
\ee
The form of the equality \eqref{sec_03_04-11} shows that the additional EMT $\bfsema$ can be defined as
\be\label{sec_03_04-15}
\boxed{
\sema\mu\nu \Def - \lp \sna\eta \belaud\mu\eta\nu + \bfrac12 \belaud\veps\eta\nu \tor\mu\veps\eta + \belaud\mu\bet\alp \tor\alp\bet\nu \rp.
}
\ee
Then, the sums $(\bffsem + \bfmsem + \bftsem + \bfsema)$ and $(\bffspi + \bfmspi + \bftspi + \bfspia)$ have to be considered  as the total canonical EMT and ST. In order to distinguish these quantities from $\bfsem$ \eqref{sec_03_04-06} and $\bfspi$ \eqref{sec_03_04-12} we call them as \emph{modified total canonical EMT and ST} and denote through $\bfsemi$ and $\bfspii$. Thus,
\be\label{sec_03_04-13}
\boxed{
\semi\mu\nu \Def \fsem\mu\nu + \msem\mu\nu + \tsem\mu\nu + \sema\mu\nu = \sem\mu\nu + \sema\mu\nu;
}
\ee
\be\label{sec_03_04-14}
\boxed{
\spii\pi\rho\sig \Def \fspi\pi\rho\sig + \mspi\pi\rho\sig + \tspi\pi\rho\sig + \spia\pi\rho\sig = \spi\pi\rho\sig + \spia\pi\rho\sig.
}
\ee
In the terms of the modified canonical tensors the identity \eqref{sec_03_04-11} is rewritten as
\be
\boxed{
\sna\mu \semi\mu\nu \eq -\semi\mu\lam \tor\lam\mu\nu + \bfrac12 \spii\pi\rho\sig \curud\rho\sig\pi\nu + \DIDffiA \na\nu \ffiA.
}
\ee
A correspondent equation of balance has the form:
\be
\boxed{
\sna\mu \semi\mu\nu = -\semi\mu\lam \tor\lam\mu\nu + \spii\pi\rho\sig \curud\rho\sig\pi\nu \quad \mbox{(on the $\bfffi$-equations)}.
}
\ee
This equation has exactly the form of the equation \eqref{sec_03_00-02}, also it takes a place only, when the equations of motion of the \emph{material} fields $\bfffi$ hold.

It is interesting to note also that, in fact, a disturbance of the $\bfffi$ -- $\bftor$ similarity occurs not because of the geometrical character of the torsion field, but as a result of non-minimality of $\bftor$-coupling (see the formulae \eqref{sec_03_04-26}, \eqref{sec_03_04-24}, \eqref{sec_03_04-15}). The non-minimal $\bftor$-interaction, on the one hand, brings a formal $\bfffi$ -- $\bftor$ similarity, on the other hand, at the same time it destroys the similarity by the terms $\bfsema$ and $\bfspia$. Note that in the case of minimal $\bftor$-coupling a contribution into the ST and EMT from the torsion as a physical field is absent at all.

Using the formulae \eqref{sec_03_04-13}, \eqref{sec_03_04-14}, \eqref{sec_03_04-15}, \eqref{sec_03_04-06}, \eqref{sec_03_04-12}, \eqref{sec_03_04-16}, it is easy to establish that the symmetrized EMT
\be\label{sec_03_04-17}
\boxed{
\sems\mu\nu \Def \semi\mu\nu + \lp \sna\eta \beliud\mu\eta\nu + \bfrac12 \beliud\veps\eta\nu \tor\mu\veps\eta + \beliud\mu\bet\alp \tor\alp\bet\nu \rp,
}
\ee
constructed with the use of the modified canonical EMT $\bfsemi$ and $\bfspii$ in the same manner as the symmetrized EMT \eqref{sec_03_04-16} coincides with $\bfsems$ \eqref{sec_03_04-16}, although the last has been constructed thorough the canonical EMT  $\bfsem$ \eqref{sec_03_04-06} and $\bfspi$ \eqref{sec_03_04-12}. Notice that $\bfbeli = 0$ on the $\bftor$-equations  (see the formulae \eqref{sec_03_04-27} and \eqref{sec_03_04-14}) and, hence, according to \eqref{sec_03_04-17}, $\bfsems$ and $\bfsemi$ are equal:
\be
\boxed{
\semi\mu\nu = \sems\mu\nu \qquad\qquad \mbox{(on the $\bftor$-equations)}.
}
\ee
%

%
\section{Summary}\label{sec_03_05-00}
In the present paper, the expressions for the canonical energy momentum tensor (EMT), spin tensor (ST), and the Belinfante symmetrized EMT have been constructed in the case, when the Lagrangian has the form
\bse
\lag = \lag (\bfmet, \bfcur;\; \bftor, \bfnator,\bfnanator;\; \bfffi, \bfnaffi, \bfnanaffi) \Def
\lag (\bfmet, \bfcur;\; \bfjfi, \bfnajfi, \bfnanajfi),
\ese
that is it contains the higher (second) covariant derivatives of the material fields $\bfffi$ as well as the non-minimal coupling both with the metric field $\bfmet$ and with the torsion field $\bftor$. It has been shown that in the presence of the higher derivatives the standard Noether procedure is ambiguous, whereas in the presence of the non-minimal $\bfmet$- or $\bftor$-coupling it is generally inapplicable (for details, see the discussion in the Sec. \ref{sec_03_00-00}). Therefore the canonical EMT and ST have been determined by the requirement that they must satisfy the standard equation of balance
\bse
\sna\mu \sem\mu\nu = -\sem\mu\lam \tor\lam\mu\nu + \frac12 \spi\pi\rho\sig \curud\rho\sig\pi\nu \qquad \mbox{(on the $\bfffi$-equations)}.
\ese
This equation was obtained earlier for the much more restricted class of field theories in Refs. \cite{Trautman_1972_b,Hehl_1973,Hehl_1974,Hehl_Heyde_Kerlick_Nester_1976,Trautman_2006}. The final most general expressions for the (modified) canonical EMT $\bfsemi$ and ST $\bfspii$ have, respectively, the form:
\bse
\ba{rl}
\semi\mu\nu & = \lf \lag \kro\mu\nu - \DIDnajfiA\mu \na\nu \jfiA - \dlagdnanajfiA\mu\lam \na\nu \na\lam \jfiA \rf\\
& + \lf -\Gu\alp\bet\gam\mu \curd\alp\bet\gam\nu \rf\\
& + \lf - \lp \sna\eta \belaud\mu\eta\nu + \bfrac12 \belaud\veps\eta\nu \tor\mu\veps\eta + \belaud\mu\bet\alp \tor\alp\bet\nu \rp \rf;
\ea
\ese
\bse
\ba{rl}
\spii\pi\rho\sig & = \lf 2 \DIDnajfiA\pi \DbrfdAB{[\rho}{\sig]} \jfiB \rd\\
& \quad \ld - 2 \dlagdnanajfiA\pi\lam \lb \met\lam{[\rho} \na{\sig]} \jfiA - \DbrfdAB{[\rho}{\sig]} \na\lam \jfiB \rb \rf\\
 & + \lf (-2)\lp \sna\eta \Gd\rho\sig\pi\eta + \frac12 \Gd\rho\sig\veps\eta \tor\pi\veps\eta \rp \rf + \lf \spia\pi\rho\sig \rf.
\ea
\ese
The \emph{nontrivial manifestly generally covariant} generalization of the Belinfante symmetrization procedure, suitable for an arbitrary Riemann-Cartan space, has been found. The correspondent symmetrized EMT $\bfsems$ has the form
\bse
\sems\mu\nu \Def \semi\mu\nu + \lp \sna\eta \beliud\mu\eta\nu + \bfrac12 \beliud\veps\eta\nu \tor\mu\veps\eta + \beliud\mu\bet\alp \tor\alp\bet\nu \rp
\ese
and satisfies the standard equation of balance
\bse
\sna\mu \sems\mu\nu = -\sems\mu\lam \tor\lam\mu\nu \qquad \mbox{(on the $\bfjfi$-equations)}.
\ese

%
\appendix
%

%
\section{The Condition for a Lagrangian to be a Scalar}\label{app_03_01-00}

Let a Lagrangian
\be\label{app_03_01-10}
\lag = \lag (\bfmet, \bfcur;\; \bftor, \bfnator,\bfnanator;\; \bfffi, \bfnaffi, \bfnanaffi)
\ee
be a generally covariant scalar. To reduce the formulae let us unite temporarily the fields $\bftor = \{ \tor\alp\bet\gam \}$ and $\bfffi = \{ \ffiA \}$ into the unique set $\bfjfi = \{ \jfiA \}$:
\bse
\bftor,\; \bfffi\quad \ra \quad \bfjfi = \{ \bftor, \bfffi \}.
\ese
Then
\bse
\lag = \lag (\bfmet, \bfcur;\; \bfjfi, \bfnajfi, \bfnanajfi).
\ese
In accordance with the definition of the scalar its total variation $\bdel\lag$ induced by an infinitesimal diffeomorphism
\be\label{app_03_01-11}
x^\mu \quad \ra \quad x'^\mu = x^\mu + \dx\mu (x),
\ee
is equal to zero:
\bse
\bdel\lag \Def \lag'(x') - \lag (x) = 0.
\ese
Taking into account the connection between the total $\bdel$ and the functional $\del$ variations
\be\label{app_03_01-06}
\bdel = \dx\lam \pa\lam + \del,
\ee
we find
\be\label{app_03_01-09}
\dx\lam \pa\lam \lag + \dlag = 0.
\ee
Let us compute $\dlag$. It is evidently
\be\label{app_03_01-12}
\ba{l}
\dlag = \lf \dslagdmet\alp\bet \dmet\alp\bet + \dlagdcur\alp\bet\gam\del \dcur\alp\bet\gam\del \rf + \lf \dslagdjfiA \djfiA + \dlagdnajfiA\kap \del \lp \na\kap \jfiA \rp \rd\\
\ld + \dlagdnanajfiA\kap\veps \del \lp \na\kap \na\veps \jfiA \rp \rf = \lf \dslagdmet\alp\bet \lb -\dx\lam \pa\lam \met\alp\bet + \bdel\met\alp\bet \rb + \dlagdcur\alp\bet\gam\del \rd\\
\ld \times \lb -\dx\lam \pa\lam \cur\alp\bet\gam\del + \bdel \cur\alp\bet\gam\del \rb \rf + \lf \dslagdjfiA \lb -\dx\lam \pa\lam \jfiA + \bdel\jfiA \rb + \dlagdnajfiA\kap \rd\\
\ld \times \lb -\dx\lam \pa\lam \lp \na\kap \jfiA \rp + \bdel \lp \na\kap \jfiA \rp \rb + \dlagdnanajfiA\kap\veps \lb -\dx\lam \pa\lam \lp \na\kap \na\veps \jfiA \rp + \bdel \lp \na\kap \na\veps \jfiA \rp \rb \rf,
\ea
\ee
where at the second step we used the connection between the total and the functional variations \eqref{app_03_01-06}. In the last formula $\pa{}^*\lag/\pa{}\met\bet\gam$ means \emph{explicit} derivative with respect to $\met\bet\gam$, that is the differentiation is provided only with respect $\met\bet\gam$, which are not included in $\bfcur$ and $\bfna$; analogously, $\pa{}^*\lag/\pa{}\jfiA$ means differentiation only with respect to $\jfiA$, which is not included in $\bfnajfi$ and $\bfnanajfi$. Note now that
\be\label{app_03_02-05}
\ba{l}
\lf \dslagdmet\alp\bet \pa\lam \met\alp\bet + \dlagdcur\alp\bet\gam\del \pa\lam \cur\alp\bet\gam\del \rf + \lf \dslagdjfiA \pa\lam \jfiA + \dlagdnajfiA\kap \pa\lam \lp \na\kap \jfiA \rp \rd\\
\ld + \dlagdnanajfiA\kap\veps \pa\lam \lp \na\kap \na\veps \jfiA \rp \rf = \pa\lam \lag.
\ea
\ee
Next, take into account tensorial nature of the quantities $\{ \met\alp\bet \}$, $\{ \cur\alp\bet\gam\del \}$, $\{ \jfiA \}$, $\{ \na\kap \jfiA \}$, $\{ \na\kap \na\veps \jfiA \}$. Then, by definition, for the infinitesimal diffeomorphisms \eqref{app_03_01-11} we have:
\begin{empheq}[left=\empheqlbrace]{align}
\bdel\met\alp\bet & = \Dbrm\sig\rho\alp\bet\eta\zet \met\eta\zet \times \pa\sig \dx\rho;\label{app_03_01-07}\\
\bdel\cur\alp\bet\gam\del & = \Dbrc\sig\rho\alp\bet\gam\del\eta\zet\vphi^\xi \cur\eta\zet\vphi\xi \times \pa\sig \dx\rho;\\
\bdel\jfiA & = \DbrjAB\sig\rho \jfiB \times \pa\sig \dx\rho;\\
\bdel \lp \na\kap \jfiA \rp & = \lb \DbrjAB\sig\rho \na\kap \jfiB - \lp \kro\sig\kap \kro\pi\rho \rp \na\pi \jfiA \rb \times \pa\sig \dx\rho;\\
\bdel \lp \na\kap \na\veps \jfiA \rp & = \lb \DbrjAB\sig\rho \na\kap \na\veps \jfiB - \lp \kro\sig\kap \kro\pi\rho \rp \na\pi \na\veps \jfiA - \lp \kro\sig\veps \kro\pi\rho \rp \na\kap \na\pi \jfiA \rb \pa\sig \dx\rho,\label{app_03_01-08}
\end{empheq}
where $\{ \DbrjAB\sig\rho \}$ are the Belinfante-Rosenfeld symbols (see, for example, Ref.~\cite{Lompay_Petrov_2013_b}). Using in the right hand side of the formula \eqref{app_03_01-12} the  formulae \eqref{app_03_02-05} -- \eqref{app_03_01-08}, we find
\bse
\ba{l}
\dlag = - \lp \pa\lam \lag \rp \times \dx\lam + \lp \lf \dslagdmet\alp\bet \Dbrm\sig\rho\alp\bet\eta\xi \met\eta\xi + \dlagdcur\alp\bet\gam\del \Dbrc\sig\rho\alp\bet\gam\del\eta\zet\vphi^\xi \cur\eta\zet\vphi\xi \rf \rd\\
\ld + \lf \dslagdjfiA \DbrjAB\sig\rho \jfiB + \dlagdnajfiA\kap \lb \DbrjAB\sig\rho \na\kap \jfiB - \kro\sig\kap \na\rho \jfiA \rb \rd \rd\\
\ld \ld + \dlagdnanajfiA\kap\veps \lb \DbrjAB\sig\rho \na\kap \na\veps \jfiB - \kro\sig\kap \na\rho \na\veps \jfiA - \kro\sig\veps \na\kap \na\rho \jfiA \rb \rf \rp \times \pa\sig \dx\rho.
\ea
\ese
Substituting this expression into the formula \eqref{app_03_01-09} and taking into account the arbitrariness of the vector field $\{ \dx\mu(x) \}$, we obtain
\be\label{app_03_02-07}
\boxed{
\ba{l}
\lf \dslagdmet\alp\bet \Dbrm\sig\rho\alp\bet\eta\xi \met\eta\xi + \dlagdcur\alp\bet\gam\del \Dbrc\sig\rho\alp\bet\gam\del\eta\zet\vphi^\xi \cur\eta\zet\vphi\xi \rf\\
\ld + \lf \dslagdjfiA \DbrjAB\sig\rho \jfiB + \dlagdnajfiA\kap \lb \DbrjAB\sig\rho \na\kap \jfiB - \kro\sig\kap \na\rho \jfiA \rb \rd \rd\\
\ld + \dlagdnanajfiA\kap\veps \lb \DbrjAB\sig\rho \na\kap \na\veps \jfiB - \kro\sig\kap \na\rho \na\veps \jfiA - \kro\sig\veps \na\kap \na\rho \jfiA \rb \rf \eq 0.
\ea
}
\ee

%
\section{The Calculation of $\na\nu \lag$}\label{app_03_02-00}
For the transformations presented in the main text of the paper one needs the explicit expression for the $\na\nu \lag$. Let us calculate it. Because $\lag$ is a generally covariant scalar of the type \eqref{app_03_01-10} one has
\be\label{app_03_02-06}
\na\nu \lag = \pa\nu \lag = \mbox{L.H.S. of the eq. } \eqref{app_03_02-05}.
\ee
 Using the expressions for the covariant derivatives $\{ \na\nu \met\alp\bet \}$, $\{ \na\nu \cur\alp\bet\gam\del \}$, $\{ \na\nu \jfiA \}$, $\{ \na\nu \na\kap \jfiA \}$, $\{ \na\nu \na\kap \na\veps \jfiA \}$
\begin{empheq}[left=\empheqlbrace]{align}
\na\nu \met\alp\bet & = \pa\nu \met\alp\bet + \con\rho\sig\nu \Dbrm\sig\rho\alp\bet\eta\zet \met\eta\zet;\\
\na\nu \cur\alp\bet\gam\del & = \pa\nu \cur\alp\bet\gam\del + \con\rho\sig\nu \Dbrc\sig\rho\alp\bet\gam\del\eta\zeta\vphi^\xi\; \cur\eta\zeta\vphi\xi;\\
\na\nu \jfiA & = \pa\nu \jfiA + \con\rho\sig\nu \DbrjAB\sig\rho \jfiB;\\
\na\nu \lp \na\kap \jfiA \rp & = \pa\nu \lp \na\kap \jfiA \rp + \con\rho\sig\nu \lb \DbrjAB\sig\rho \na\kap\jfiB - \kro\sig\kap \na\rho \jfiA \rb;\\
\na\nu \lp \na\kap \na\veps \jfiA \rp & = \pa\nu \lp \na\kap \na\veps \jfiA \rp\nonumber\\
 & + \con\rho\sig\nu \lb \DbrjAB\sig\rho \na\kap \na\veps \jfiB - \kro\sig\kap \na\rho \na\veps \jfiA - \kro\sig\veps \na\kap \na\rho \jfiA \rb,
\end{empheq}
 we find the partial derivatives $\{ \pa\nu \met\alp\bet \}$, $\{ \pa\nu \cur\alp\bet\gam\del \}$, $\{ \pa\nu \jfiA \}$, $\{ \pa\nu (\na\kap \jfiA) \}$, $\{ \pa\nu (\na\kap \na\veps \jfiA) \}$ and substitute them into the formula \eqref{app_03_02-06}. After a rearrangement of items we obtain
\be
\ba{l}
\na\nu \lag = \lp \lf \dslagdmet\alp\bet \na\nu \met\alp\bet + \dlagdcur\alp\bet\gam\del \na\nu \cur\alp\bet\gam\del \rf + \lf \dslagdjfiA \na\nu \jfiA + \dlagdnajfiA\kap \na\nu \lp \na\kap \jfiA \rp \rd \rd\\
\ld \ld + \dlagdnanajfiA\kap\veps \na\nu \lp \na\kap \na\veps \jfiA \rp \rf \rp - \con\rho\sig\nu \times \lp \mbox{ L.H.S. of the eq. } \eqref{app_03_02-07} \rp.
\ea
\ee
Taking into account in this relationship the identity \eqref{app_03_02-07} and metric-compatible condition $\na\nu \met\alp\bet = 0$, we find the search expression:
\be
\boxed{
\ba{rl}
\na\nu \lag & = \lf \dlagdcur\alp\bet\gam\del \na\nu \cur\alp\bet\gam\del \rf\\
& + \lf \dslagdjfiA \na\nu \jfiA + \dlagdnajfiA\kap \na\nu \na\kap \jfiA + \dlagdnanajfiA\kap\veps \na\nu \na\kap \na\veps \jfiA \rf.
\ea
}
\ee
For the cases of interest this expression takes the following forms:
\bn
\item The case of minimal coupling, $\lag = \lag (\bfmet;\; \bfffi, \bfnaffi, \bfnanaffi)$,
\be\label{sec_03_02-07}
\na\nu \lag = \lf \dslagdffiA \na\nu \ffiA + \dlagdnaffiA\kap \na\nu \na\kap \ffiA + \dlagdnanaffiA\kap\veps \na\nu \na\kap \na\veps \ffiA \rf;
\ee
\item The case of non-minimal $\bfmet$-coupling, $\lag = \lag (\bfmet, \bfcur;\; \bfffi, \bfnaffi, \bfnanaffi)$,
\be\label{sec_03_02-08}
\ba{rl}
\na\nu \lag & = \lf \dlagdcur\alp\bet\gam\del \na\nu \cur\alp\bet\gam\del \rf\\
& + \lf \dslagdffiA \na\nu \ffiA + \dlagdnaffiA\kap \na\nu \na\kap \ffiA + \dlagdnanaffiA\kap\veps \na\nu \na\kap \na\veps \ffiA \rf;
\ea
\ee
\item The case of non-minimal $\bfmet$- and $\bftor$-coupling, $\lag = \lag (\bfmet, \bfcur;\; \bftor, \bfnator,\bfnanator;\; \bfffi, \bfnaffi, \bfnanaffi)$,
\be\label{sec_03_04-20}
\ba{rl}
\na\nu \lag & = \lf \dlagdcur\alp\bet\gam\del \na\nu \cur\alp\bet\gam\del \rf\\
& + \lf \dslagdffiA \na\nu \ffiA + \dlagdnaffiA\kap \na\nu \na\kap \ffiA + \dlagdnanaffiA\kap\veps \na\nu \na\kap \na\veps \ffiA \rf\\
 & + \lf \dslagdtor\alp\bet\gam \na\nu \tor\alp\bet\gam + \dlagdnator\kap\alp\bet\gam \na\nu \na\kap \tor\alp\bet\gam + \dlagdnanator\kap\veps\alp\bet\gam \na\nu \na\kap \na\veps \tor\alp\bet\gam \rf.
\ea
\ee

\en

%
\section{The Transformation of the Expression $\lp \frac12 \Gu\alp\bet\gam\del \na\nu \curd\alp\bet\gam\del \rp$}\label{app_03_03-00}
Transform the expression $\frac12 \Gu\alp\bet\gam\del \na\nu \curd\alp\bet\gam\del$ as follows.
Substituting the Ricci identity in the form
\bse
\na\nu \curd\alp\bet\gam\del \eq - \lp \na\gam \curd\alp\bet\del\nu + \na\del \curd\alp\bet\nu\gam + \curd\alp\bet\veps\nu \tor\veps\gam\del + \curd\alp\bet\veps\gam \tor\veps\del\nu + \curd\alp\bet\veps\del \tor\veps\nu\gam \rp,
\ese
one obtains
\bse
\frac12 \Gu\alp\bet\gam\del \na\nu \curd\alp\bet\gam\del = -\Gu\alp\bet\gam\del \na\gam \curd\alp\bet\del\nu - \lp \Gu\alp\bet\gam\del \curd\alp\bet\veps\gam \rp \tor\veps\del\nu - \frac12 \lp \Gu\alp\bet\gam\del \tor\veps\gam\del \rp \curd\alp\bet\veps\nu.
\ese
Differentiating by parts the first item in the right hand side one finds
\bse
\boxed{
\ba{rl}
\bfrac12 \Gu\alp\bet\gam\del \na\nu \curd\alp\bet\gam\del & = \lb \sna\mu \lp \Gu\alp\bet\gam\mu \curd\alp\bet\gam\nu \rp + \lp \Gu\alp\bet\gam\mu \curd\alp\bet\gam\lam \rp \tor\lam\mu\nu \rb\\
 & + \bfrac12 \lb (-2)\lp \sna\eta \Gu\rho\sig\pi\eta + \bfrac12 \Gu\rho\sig\veps\eta \tor\pi\veps\eta \rp \rb \curd\rho\sig\pi\nu.
\ea
}
\ese
%

%
\section{The Transformation of the Expression $\lp \frac{-1}{2} \belud\gam\bet\alp \na\nu \tor\alp\bet\gam \rp$}\label{app_03_04-00}
Let $\{ \belu\gam\bet\alp \} \Def \{ \Du\gam\bet\alp\pi\rho\sig
\spiu\pi\rho\sig \}$, where $\{ \spiu\pi{[\rho}{\sig]} =
\spiu\pi\rho\sig \}$, be an \emph{arbitrary} tensor with such a symmetry.
Then $\belu{[\gam}{\bet]}\alp = \belu\gam\bet\alp$. Based on this,
transform the expression $\lp \frac{-1}{2} \belud\gam\bet\alp \na\nu
\tor\alp\bet\gam \rp$ as follows.
\bn
\item Substituting the Ricci identity in the form
\bse \ba{rl}
\na\nu \tor\alp\bet\gam & \eq \cur\alp\nu\bet\gam + \cur\alp\bet\gam\nu + \cur\alp\gam\nu\bet\\
& - \lp \na\bet \tor\alp\gam\nu + \na\gam \tor\alp\nu\bet + \tor\alp\lam\nu \tor\lam\bet\gam + \tor\alp\lam\bet \tor\lam\gam\nu + \tor\alp\lam\gam \tor\lam\nu\bet \rp\\
& = \cur\alp\nu\bet\gam + 2\cur\alp{[\bet}{\gam]}\nu - 2\na{[\bet}
\tor\alp{\gam]}\nu -  \tor\alp\lam\nu \tor\lam\bet\gam -
2\tor\alp\lam{[\bet} \tor\lam{\gam]}\nu, \ea \ese
one obtains
\be\label{app_03_04-01}
\ba{rl}
-\bfrac12 \belud\gam\bet\alp \na\nu \tor\alp\bet\gam & \eq -\bfrac12 \belud\gam\bet\alp \cur\alp\nu\bet\gam - \belud\gam\bet\alp \cur\alp\bet\gam\nu + \belud\gam\bet\alp \na\bet \tor\alp\gam\nu\\
& + \bfrac12 \belud\gam\bet\alp \tor\alp\lam\nu \tor\lam\bet\gam +
\belud\gam\bet\alp \tor\alp\lam\bet \tor\lam\gam\nu.
\ea
\ee
\item \label{app_03_04-02} Turn to the first term on the
right  hand side of \eqref{app_03_04-01}. Then, recall the identity
(C2) in the Appendix C.1 of the Ref.~\cite{Lompay_Petrov_2013_a}:
\bse \sna\mu \lb \sna\eta \potb\nu\mu\eta +
\bfrac{1}{2}\potb\nu\rho\sig \tor\mu\rho\sig \rb \eq -\bfrac{1}{2}
\cur\lam\nu\rho\sig \potb\lam\rho\sig, \ese
change here $\potb\nu\mu\eta = \belud\mu\eta\nu$ and obtain for this
term:
\bse -\bfrac12 \belud\gam\bet\alp \cur\alp\nu\bet\gam = -\sna\mu \lb
\sna\eta \belud\mu\eta\nu + \bfrac12 \belud\veps\eta\nu
\tor\mu\veps\eta\rb. \ese
\item The second term on the right  hand side of \eqref{app_03_04-01} is equal to
\bse \ba{l}
-\belud\gam\bet\alp \cur\alp\bet\gam\nu = -\belu\gam\bet\alp \curd\alp\bet\gam\nu = -\Du\gam\bet\alp\pi\rho\sig \spiu\pi\rho\sig \curd\alp\bet\gam\nu\\
\quad = -\bfrac12 \lp \spiu\bet\gam\alp + \spiu\alp\gam\bet - \spiu\gam\bet\alp \rp \curd\alp\bet\gam\nu\\
\quad = \lp \spiu{(\alp}{\bet)}\gam - \frac12 \spiu\gam\alp\bet \rp
\curd\alp\bet\gam\nu = -\bfrac12 \spiu\pi\rho\sig
\curd\rho\sig\pi\nu; \ea \ese
\item Using the differentiation by part in the third term on the
right  hand side of \eqref{app_03_04-01}, one finds
\bse \belud\gam\bet\alp \na\bet \tor\alp\gam\nu = -\sna\mu \lp
\belud\mu\bet\alp \tor\alp\bet\nu \rp - \lp \sna\eta
\belud\mu\eta\lam \rp \tor\lam\mu\nu; \ese

\item\label{app_03_04-03} At last, one rewrites fourth and fifth terms on the
right  hand side of \eqref{app_03_04-01}, respectively, as
\bse \frac12 \belud\gam\bet\alp \tor\alp\lam\nu \tor\lam\bet\gam =
-\frac12 \lp \belud\veps\eta\lam \tor\mu\veps\eta \rp
\tor\lam\mu\nu \ese
and
\bse \belud\gam\bet\alp \tor\alp\lam\bet \tor\lam\gam\nu = - \lp
\belud\mu\bet\alp \tor\alp\bet\lam \rp \tor\lam\mu\nu. \ese
\en
Combining the results of the points \ref{app_03_04-02} --
\ref{app_03_04-03} in the formula \eqref{app_03_04-01}, one obtains
the search identity:
\be\label{sec_03_04-08}
\boxed{
\ba{rl}
-\bfrac12 \belud\gam\bet\alp \na\nu \tor\alp\bet\gam & \eq - \sna\mu \lb \sna\eta \belud\mu\eta\nu + \bfrac12 \belud\veps\eta\nu \tor\mu\veps\eta + \belud\mu\bet\alp \tor\alp\bet\nu \rb \\
 & - \lb \sna\eta \belud\mu\eta\lam + \bfrac12 \belud\veps\eta\lam \tor\mu\veps\eta + \belud\mu\bet\alp \tor\alp\bet\lam \rb \tor\lam\mu\nu - \bfrac12 \spi\pi\rho\sig \curud\rho\sig\pi\nu.
\ea
}
\ee
%

%

%
\bibliographystyle{spmpsci}      

\bibliography{Lompay_2013_a_arXiv_Bibliography}

\end{document}